\documentclass[twocolumn]{article}

\usepackage[english]{babel}

\usepackage[left=2.0cm, right=2.0cm, top=2cm, bottom=2cm]{geometry}
\usepackage{subfiles} % For handling multi-file projects
\usepackage{cuted}    % For the 'strip' environment (full-width text in 2-col)

\usepackage{amsmath}
\usepackage{amsthm}
\usepackage{array}

\usepackage{graphicx}
\usepackage{flafter}  % Ensures floats do not appear before their definition

\usepackage[colorlinks=true, allcolors=blue]{hyperref}
\usepackage{authblk}  % For formatting author affiliations

\usepackage{fancyhdr}

\usepackage[backend=biber, style=numeric, sorting=none]{biblatex}
\fancypagestyle{plain}{
  \fancyhf{}
  \fancyhead[L]{\textit{Preprint version}}
  \fancyhead[R]{\today}
  \fancyfoot[C]{\thepage}
  
}

\title{Segregation Before Polarization: \\ How Recommendation Strategies Shape Echo Chamber Pathways}

\author[1]{Junning Zhao}
\author[2]{Kazutoshi Sasahara}
\author[1]{Yu Chen}

\affil[1]{
  Department of Human and Engineered Environmental Studies, Graduate School of Frontier Sciences, The University of Tokyo,
  5-1-5 Kashiwanoha, Kashiwa, Chiba 277-8563, Japan
}
\affil[2]{
  Department of Innovation Science, School of Environment and Society, Institute of Science Tokyo, 
  3-3-6 Shibaura, Minato, Tokyo 108-0023, Japan
}

\date{} % Empty date to suppress default date insertion

\begin{document}

\maketitle

% --- Main Text Reference Section ---
\begin{refsection}

    \begin{abstract}
        Social media platforms facilitate echo chambers through feedback loops between user preferences and recommendation algorithms. While algorithmic homogeneity is well-documented, the distinct evolutionary \emph{pathways} driven by content-based versus link-based recommendations remain unclear. Using an extended dynamic Bounded Confidence Model (BCM), we show that content-based algorithms—unlike their link-based counterparts—steer social networks toward a segregation-before-polarization (SbP) pathway. Along this trajectory, structural segregation precedes opinion divergence, accelerating individual isolation while delaying but ultimately intensifying collective polarization. Furthermore, we reveal that reposting appears connective by circulating content beyond direct follow links, yet it simultaneously reinforces echo chambers because it amplifies small, latent opinion differences that would otherwise remain inconsequential. These findings suggest that mitigating polarization requires stage-dependent algorithmic interventions, shifting from content-centric to structure-centric strategies as networks evolve.

\paragraph{Significance}

Social media platforms are plagued by algorithmically-driven echo chambers that enhance polarization, but the specific ways different recommendation strategies contribute to this problem are not well understood. We use a novel computational model to disentangle the effects of content-based versus link-based recommendations on the evolutionary path of online social networks. We demonstrate that content-based algorithms, now common on major platforms, create a segregation-before-polarization pathway, accelerating an individual's isolation and leading to more stable, intensely polarized groups. Our findings also reveal the paradoxical role of information sharing (reposting) in amplifying polarization and provide a theoretical background for stage-dependent interventions, suggesting that platforms could dynamically adjust algorithms to mitigate polarization without resorting to censorship.
    \end{abstract}

\section{Introduction}\label{sec:introduction}

Social media has become a primary platform for obtaining information and exchanging opinions on major news and events. Beyond its role in significant political and economic moments , such as elections or financial crises, it allows users to share their personal lives in rich, multimedia formats. However, a growing body of research indicates that the rapid information transmission on these platforms can lead to detrimental outcomes, most notably selective exposure and political polarization \cite{bail2018exposure, mok2023echo, tornberg2022digital}. Many studies demonstrate that social media's characteristics can foster highly polarized and segregated social structures, leading to distinctive network topologies, such as networks with high clustering coefficients or fragmentation into sparsely interconnected communities \cite{sasahara2021social, santos2021link, tornberg2022digital, piccoli2018sparse}.

The dynamics of social interaction are broadly governed by two processes: social influence and social selection. The former refers to the situation where individuals' thoughts and behaviors are shaped by others, while the latter refers to the connections of individuals that are formed based on shared characteristics.
Social media platforms facilitate both and amplify them dramatically. The ease of following or unfollowing other users, for instance, results in a rapid evolution of network structure. Moreover, this dynamic environment fosters \emph{selective exposure}, the tendency for users to favor content that reinforces their existing views while avoiding opposing ones \cite{alatawi2021survey}. Users can actively manage, on the online platforms, their information feeds by disconnecting from those with different viewpoints \cite{bozdag2020managing}. Empirical evidence, which ranged from political polarization in the U.S. to public health debates like vaccination during the COVID-19 pandemic, confirms that selective exposure is a significant consequence. For example, exposure to opposing political views on Twitter has been shown to exacerbate polarization rather than mitigate it \cite{bail2018exposure, grover2019polarization}. Some models even suggest that the intrinsic nature of social media, with its rapid network rewiring, can accelerate social fragmentation and partisan hostility, even with a random content exposure \cite{sasahara2021social, tornberg2022digital}.

Most online platforms, including social media, rely on recommendation algorithms to filter vast amounts of information and suggest relevant items—such as posts, articles, or products—to their users. These systems are typically \emph{personalized}, by analyzing and taking advantage from user preferences to maximize engagement and platform loyalty \cite{li2024recent, mansoury2020feedback}. While undeniably useful, these algorithms create a closed feedback loop: The system recommends content based on a user's current preferences; the user then responds with the recommendation; this interaction will refine preferences of the user, which in turn shapes future recommendations. 
Previous studies have confirmed that such kind of feedback loop can lead to preference drift\cite{kalimeris2021preference, rossi2022closed, schmit2018human}, increased homogeneity\cite{chaney2018algorithmic, ekstrand2018all}, and reduced content diversity \cite{jiang2019degenerate, anderson2020algorithmic, jannach2017efficient, mansoury2020feedback}.

Note that this feedback loop is further intensified by two unique characteristics of social media: 
First, users are both the producers and consumers of content; and second, the content production cycle is exceptionally fast. This high-speed environment suggests that the dynamic effects of recommendation algorithms may extend beyond simple preference reinforcement, potentially shaping the entire evolutionary path of the social system in more complex ways. For instance, information sharing behavior in the social media (such as reposting, retweeting, etc.) would amplify and filter the result produced by the recommendation algorithm, so that the recommendation acts as a secondary dissemination mechanism of algorithmic decision-making in social networks \cite{huszar2022algorithmic, buntain2021youtube, dearruda2024echo, oliveira2026mechanistic}.

Social media platforms employ diverse recommendation strategies. 
A common approach is \emph{link-based} recommendation, which suggests connections based on network proximity (e.g., ``friends of friends''). Studies have shown that such systems can enhance social fragmentation and polarization \cite{santos2021link, morales2021auditing}. 
A notable trend in recommendation strategies is the increasing adoption of \emph{content-based} recommendation algorithms, which analyze the substance of posts (e.g., topic, sentiment) in order to make suggestions \cite{grewal2018evolution, yang2022personality, ying2022recommendation}.
However, the contribution of content-based recommendation algorithms to the formation of echo chambers has not been clarified yet.

% A closer look at real-world platforms reveals that echo chambers do not form uniformly. 
An echo chamber usually refers to a social structure in which individuals are primarily exposed to information, opinions, and perspectives that reinforce their existing beliefs\cite{alatawi2021survey}.
The formation and evolution of echo chambers differ across platforms, resulting in varied pathways\cite{morales2021no}.
On one hand, some platforms may exhibit extreme local opinion homogenization before the whole society is polarized and segregated.
For instance, YouTube’s recommendation engine has been shown to induce highly modular and homophilous network structures, creating favorable conditions for filter bubbles \cite{kaiser2020birds}. In contrast, Zhihu’s shift toward social filtering primarily led to a concentration of user interests, indicating a direct effect on preference formation \cite{liu2023daily}. 
On the other hand, some platforms are characterized by polarization between distinct camps in early stages of echo chamber formation. Twitter (X) appears to be the most distinct example of this category.
An empirical study shows that those accounts that follow Twitter's 'Who-To-Follow' recommendation have a more dense and interconnected network with a lower political homogeneity than networks established through social endorsement \cite{duskin2024echo}. 
Another paper shows that a diverse but mildly polarized society could emerge in strong biased assimilation condition\cite{morales2021auditing}.
Some research has explored strategies to mitigate these effects, such as by recommending diverse content or using control theory to de-cluster networks \cite{donkers2021dual, piccoli2018sparse, dean2022preference}. Yet many of these studies do not fully capture the interplay between content consumption and network evolution that is central to modern social media. 

These findings above suggest that the critical question is not merely \emph{if} a system will form an echo chamber, but \emph{how} and \emph{along which pathway} it evolves, because this \emph{pathway} not only determines the final structure of the echo chamber but also suggests effective interventions preventing the echo chamber. Nevertheless there is still lack of understanding on how specific algorithmic mechanisms generate the diverse paths of echo chamber formation. This critical gap leads to our primary research questions:

\textbf{(RQ1)} \emph{Comparing to link-based ones, how do content-based recommendation algorithms differentially influence the evolutionary pathways of echo chamber formation and its final state?}

\textbf{(RQ2)} \emph{How does the information sharing act as a moderator variable to amplify or suppress the aspects of echo chamber along different pathways?}

% With respect to different algorithms and societies,
\textbf{(RQ3)} \emph{What are the pathways' societal meanings in individual- and collective-level perspectives?}

% \textbf{(RQ4)} \emph{At what stages in the echo chamber formation process do content-based and link-based recommendation systems offer different intervention opportunities for mitigating polarization?}

To address these questions, we turn to the tradition of agent-based opinion dynamics modeling, particularly the Bounded Confidence Model (BCM) framework \cite{hegselmann2002opinion, deffuant2000mixing, degroot1974reaching, ren2005survey}. BCMs are effective at studying how consensus, polarization, and fragmentation emerge from simple interaction rules based on opinion proximity (social influence and selection). While classical BCMs capture the core of opinion clustering, they often abstract away the algorithmic curation and high-speed production-consumption cycles of modern social networks \cite{sasahara2021social, sirbu2019algorithmic}.

In this work, we build on this tradition by extending a discrete BCM to incorporate algorithmic curation---both link-based and content-based---on a dynamic follower network. This allows us to systematically examine not only the end states but also the diverse pathways leading to echo chamber formation. We propose several metrics for opinion polarization and homophily to quantify the effects of two idealized recommendation systems that mimic content-based and link-based strategies. We perform extensive simulations across virtual societies with varying tendencies for social influence, network rewiring, and information sharing (reposting / retweeting). Our results show that while all systems eventually converge to a polarized or consensual state, the evolutionary trajectory in terms of homophily and polarization differs significantly. In a society where the effect of social influence and social selection equates, content-based recommendations shift the pathway towards a segregation-before-polarization (SbP) one. This results in the acceleration of individuals' entry into echo chambers and more intense, faster-forming, collective echo chambers. In contrast, link-based recommendation drive the system to the final state more typically via a polarization-before-segregation (PbS) pathway.

The remainder of this paper is organized as follows: 
Section \ref{sec:methods} details our proposed extended dynamic bounded-confidence model (BCM), elaborates on the mechanisms for content and link recommendations, and defines key metrics for quantifying network segregation, opinion polarization, and evolutionary pathway.
Section \ref{sec:results} presents simulation results, first analyzing the potential energy landscape of opinion evolution as well as explaining it with Landau theory, then compares the distinct echo chamber formation pathways led by different recommendation strategies, and lastly discuss how information sharing behavior acts as a moderating variable influencing individual isolation and group polarization. Section \ref{sec:discussion} discusses the theoretical and practical implications of the findings. Based on the simulation results, it proposes targeted phased algorithmic intervention strategies, identifies the limitations of this study, and outlines future research directions.

    \section{Methods}\label{sec:methods}

To investigate the interplay between algorithmic content curation and social dynamics, we first developed an agent-based model where opinion formation and network structure co-evolve. Social dynamics unfold on a directed graph $G = (V, E)$, where each node $i \in V$ represents an agent (user of the social media), and a directed edge $(i, j) \in E$ signifies that agent $i$ follows agent $j$. This structure captures the unidirectional follower relationships typical of social media platforms such as Twitter (X). Each agent has a continuous variable of opinion $x_i \in [-1, 1]$, which is updated synchronously at discrete time.

\subsection{Model Description}

At each time step, agents update their opinions based on exposure to content, potentially rewire their network connections, and generate new content.

\paragraph{Opinion Update}
Agents are exposed to content through \emph{posts}. A post is a triad $(i, t, \tau)$, where $i$ is the author's ID, $t$ is the posting time, and $\tau$ is the opinion it carries. We assume that a post's opinion $\tau$ is identical to its author's opinion at that time, i.e. $\tau = x_i(t)$.

Information from followees and from recommendations are displayed together on the timeline. Mathematically, for each agent $i$ at time $t$, it receives posts from two sets: the set of posts from its followees $A_i(t) := \{(j, t - 1, \cdot)|(i, j)\in E\}$; and the set $R_i(t)$ of fixed size $k_R$ provided by the platform's recommendation system.
Agent $i$ deems a post with opinion $\tau$ of agent $j$ or the recommendation system as \emph{concordant} if it falls within its \emph{confidence boundary} $\epsilon$, i.e., $|\tau - x_i(t)| < \epsilon$. Otherwise, the post is deemed as a \emph{discordant} one.

The sets of concordant posts from followees and recommendations are, respectively:
\begin{align*}
    N_i(t) &:= \{(\cdot, \cdot, \tau)|(\cdot, \cdot, \tau) \in A_i(t), |x_i(t) - \tau| < \epsilon\}, \\
    M_i(t) &:= \{(\cdot, \cdot, \tau)|(\cdot, \cdot, \tau) \in R_i(t), |x_i(t) - \tau| < \epsilon\}.
\end{align*}
Agent $i$ updates its opinion by averaging the opinions from all concordant posts it receives \cite{hegselmann2002opinion, sasahara2021social}:
\begin{align*}
    x_i(t+1)  &= x_i(t) \\
     &+ \frac{\alpha}{|N_i(t)|+|M_i(t)|} \sum_{(\cdot, \cdot, \tau) \in N_i(t) \cup M_i(t)} \left( \tau - x_i(t) \right),
\end{align*}
where $\alpha \ge 0$ is the \emph{influence parameter} which controls the strength of social influence. If both sets are null for certain agent, the update will be skipped and agent's opinion remains unchanged.

\paragraph{Network Rewiring}
To simulate social selection, each agent may renew its social ties. With a \emph{rewiring probability} $q$, agent $i$ randomly selects one discordant followee (i.e., agent $j$ such that $\exists \  (j, \cdot, \cdot) \in (A_i(t) - N_i(t))$ ), unfollows them, and switches to a new agent $k$ who authored a concordant recommended post (i.e., $(k, \cdot, \cdot) \in M_i(t)$). This mechanism models the tendency of users to curate their social environment toward opinion-reinforcing sources.

\paragraph{Content Generation}
Each agent may also contribute contents to the social network. At time $t$, with a \emph{repost probability} $p$, agent $i$ reposts a randomly chosen concordant post from $N_i(t) \cup M_i(t)$. With probability $1-p$, it posts a new, original post $(i, t, x_i(t))$. This new content becomes available to its followers as well as to the recommendation system at time $t + 1$. This mechanism is crucial to model the information sharing behavior.

\subsection{Recommendation Strategies}
We implement three recommendation algorithms to compare the effects of different curation logic on the emergence of echo chambers.
\begin{description}
    \item[Random] This baseline algorithm serves as a null model. It selects $k_R$ posts for each agent uniformly at random from all recommendable posts (namely, all posts not authored by the agent or its current followees).

    \item[Structure-based] This strategy mimics idealized link recommendation systems that suggest users with similar network neighborhoods. It first identifies agents that are not followed by the user, and meanwhile have the highest structural similarity (common neighbors count). The system then recommends up to $k_R$ the most recent posts from this set of agents.
    
    \item[Opinion-based] This strategy serves as an idealized content-based filter that prioritizes posts whose opinions are closest to an agent’s current stance. From a pool of candidate posts generated within a recent time window of length $k_h$, the algorithm selects $k_R$ posts that minimize the opinion distance between the target agent and the content producers.
\end{description}

\subsection{Measurement Criteria}\label{sec:measurement-criteria}
To quantify the system's collective behavior and the formation of echo chambers, we propose a set of indices that capture different facets of the phenomenon, including structural homophily, polarization, and network structure.

\paragraph{Structural Homophily Index}
Homophily, understood here as the observed tendency of network connections to align with opinion similarity, is measured as a property of the network structure. We define homophily ratio $\rho_\epsilon(t)$ and homophily index $I_h(t)$ as the normalized proportion of concordant connections in the network. It is calculated as:
\begin{align*}
    \rho_\epsilon(t) &:= \frac{1}{|V|} \sum_{i \in V}\frac{|\{j | (i, j) \in E, |x_i(t) - x_j(t)| < \epsilon \}|}{|\{j | (i, j) \in E\}|}, \\
    I_h(t) &:= \max\left(0, \frac{\rho_\epsilon(t) - \overline{\rho_\epsilon(0)}}{1 - \overline{\rho_\epsilon(0)}}\right)
\end{align*}
where $G = (V, E)$ defines the following network, and $\overline{\rho_\epsilon(0)}$ is the expected homophily ratio for a randomly initialized scenario ($\overline{\rho_\epsilon(0)} = \epsilon - \frac{\epsilon^2}{8}$ for an Erd\H{o}s-R\'enyi graph with uniformly selected opinions).
This index measures the extent to which the social network is segregated into opinion-aligned clusters, with $I_h=0$ representing a randomly connected state and $I_h \approx 1$ indicating a state of high structural homophily, i.e., agents having similar opinions are connected with each other.

\paragraph{Polarization Indices}
To measure opinion fragmentation, we analyze the distribution of opinion distances. The \emph{objective opinion distance distribution}, $f^O(\Delta x;t)$, considers all pairs of (connected or unconnected) agents in the network, reflecting global polarization. The \emph{subjective opinion distance distribution}, $f^S(\Delta x;t)$, considers only connected pairs (follower-followee), reflecting the opinion environment in an agent's local neighborhood.

We quantify the degree of polarization and local clustering by comparing these empirical distributions to reference distributions using the Jensen-Shannon divergence \cite{lin1991divergence}. 
A set of random distribution ($f^O_R, f^S_R$) and a set of fully clustered distribution ($f^O_C, f^S_C$) are employed as references. The \emph{objective polarization index} $I_p(t)$ and \emph{subjective polarization index} $I_s(t)$ are defined as:
\begin{align*}
    I_p(t) &= \frac{D_\text{JS}(f^O(\Delta x;t) \| f^O_R)}{D_\text{JS}(f^O_C \| f^O_R)}, \\
    I_s(t) &= \frac{D_\text{JS}(f^S(\Delta x;t) \| f^S_R)}{D_\text{JS}(f^S_C \| f^S_R)}.
\end{align*}
Notice that $I_p(t)$ measures the degree to which the society's opinions have split into distinct groups, while $I_s(t)$ refers to the homogeneity of opinions within an agent’s local information neighborhood, rather than polarization at the population level. A value of 0 for both indicates a random state, while a value of 1 indicates maximal global opinion separation for $I_p(t)$, and complete local opinion homogeneity for $I_s(t)$, respectively.

\paragraph{Pathway Index}
Although polarization often arises in such systems, it will eventually evolve into a segregated or consensual state via different evolutionary pathways. Specifically, it can go over a \emph{PbS} pathway (opinions polarize before the network segregates) or a \emph{SbP} pathway (the network segregates before opinions polarize). To characterize this, we introduce the \emph{pathway index} $I_w$, which measures the extent to which changes in polarization occur under already segregated network structures, by calculating the area under the trajectory on the $I_p$-$I_h$ plane:
\begin{equation*}
    I_w = \int_{t} I_h(t) {dI_p(t)}.
\end{equation*}
Large values of $I_w$ indicate SbP-like trajectories, whereas smaller values correspond to PbS-like trajectories.

\paragraph{Discussion}
Although the indices $I_h$, $I_p$ and $I_s$ all quantify forms of opinion alignment, they capture fundamentally different levels of social organization. The subjective index $I_s$ measures the homogeneity of agents’ local information environments and thus reflects the strength of echo chambers from an individual perspective. In contrast, the structural homophily index $I_h$ characterizes network-level segregation by quantifying the extent to which connections are concentrated among opinion-similar agents. Finally, the polarization index $I_p$ captures global opinion fragmentation across the entire population.

Importantly, these quantities are not equivalent. Local informational homogeneity ($I_s = 1$) does not necessarily imply global structural segregation ($I_h = 1$), nor does structural segregation alone guarantee maximal global polarization ($I_p = 1$) Only in the limiting case of perfectly segregated and internally homogeneous clusters do all three indices coincide. This separation of scales clarifies how echo chambers can emerge early at the local level, well before polarization becomes pronounced at the societal level—a key mechanism underlying the Segregation-before-Polarization (SbP) pathway identified in this study.

Building on this distinction, the pathway index $I_w$ complements these state-based measures by characterizing the temporal ordering between segregation and polarization along an evolutionary trajectory. Rather than describing a system state, $I_w$ summarizes whether changes in global polarization tend to occur after structural segregation has already developed. In this sense, $I_w$ provides a compact, continuous descriptor of polarization pathways, distinguishing segregation-before-polarization dynamics from trajectories in which opinion divergence precedes network segregation.

\section{Results}\label{sec:results}

We simulate opinion dynamics on directed Erdős–Rényi networks ($n=500$, $k_n=15$) with initial opinions uniform on $[-1,1]$, sweeping $\alpha,q\in[0.005,1]$ and $p\in[0,0.5]$ under both opinion- and structure-based recommendation schemes ($k_h\in\{0,2,6\}$, $k_r=10$) at fixed $\epsilon=0.45$ (see Supplementary Materials).

\subsection{Social Force Potential and the Divergence of Pathways}

As seen in Figs.~\ref{fig:toy-network}(a),(c), Two typical evolutionary pathways emerge: \emph{Polarization-before-Segregation} (PbS), where opinions diverge before the network structurally segregates, and \emph{Segregation-before-Polarization} (SbP), where structural segregation precedes opinion divergence.
To characterize the microscopic dynamics, we perform theoretical and numerical analyses to the results.

\subsubsection{Theoretical Analysis}\label{sssec:theoretical}

We define the \emph{normalized opinion difference} (NOD) $\Delta_n x_i(t) := \alpha^{-1}(x_i(t+1)-x_i(t))$. Since $\alpha$ acts as an inverse inertial mass (mobility), dividing by $\alpha$ isolates the effective \emph{Social Force} $F_i := \Delta_n x_i(t)$ exerted on agent $i$ by the opinion field. Integrating the mean social force $F(x)$ defines an effective potential
$$V(x) := C - \int_{-1}^{x} F(x')\,dx',$$
with $C$ chosen so $\int_{-1}^1 V(x)\,dx=0$, whose wells are the opinion attractors.
The system hence produces a \emph{self-consistent field}: the free energy of a single agent's opinion is determined by $V(x)$, yet $V(x)$ is itself generated by the collective opinion distribution. 

To estimate $F(x)$, we consider three idealized typical network states—(1) a random E-R network, (2) a consensual distribution $X\sim\mathcal{N}(0,\sigma_c^2)$, and (3) a diverged two-component distribution $X\sim\mathcal{N}(\pm0.5,\sigma_d^2)$—and compute
$$F(x) = \frac{\displaystyle\int_{x-\epsilon}^{x+\epsilon} x'\bigl(k_n f_n(x,x')+k_r f_r(x,x')\bigr)\,dx'}{\displaystyle\int_{x-\epsilon}^{x+\epsilon}\bigl(k_n f_n(x,x')+k_r f_r(x,x')\bigr)\,dx'} - x,$$
where $f_n,f_r$ are the opinion PDFs perceived from followees and recommended posts, respectively (see Supplementary Materials for the concrete computation). 

In a physical sense, this is comparative to a \emph{quench} procedure that lets the system to relax from a disordered initial state to a steady state. 
The initial landscape (Fig.~\ref{fig:toy-network} (e1)) shows a critical state with only shallow edge troughs and a flat interior.
The consensual final landscape (e2), (e3) shows the symmetrically stable phase with a single deep central well.
The diverged landscape (e4), (e5) shows the Landau-type broken-symmetry stable phase with a symmetric double well regarding the local order parameter (here, the opinion scalar $x$).

Notably, all three recommendation algorithms yield nearly identical potentials, indicating the landscape shape is primarily determined by the opinion distribution rather than the algorithm. 

\subsubsection{Simulation Analysis} 

Landau's theory of phase transitions provides a compelling phenomenological account of this symmetry breaking. However, because the potential is self-consistently generated by the very agents it governs---and further deformed by network rewiring and algorithmic feedback---theoretical analysis alone is insufficient. Direct simulation is therefore necessary.

\begin{figure*}[!htbp]
    \centering
    \includegraphics[width=0.95\textwidth]{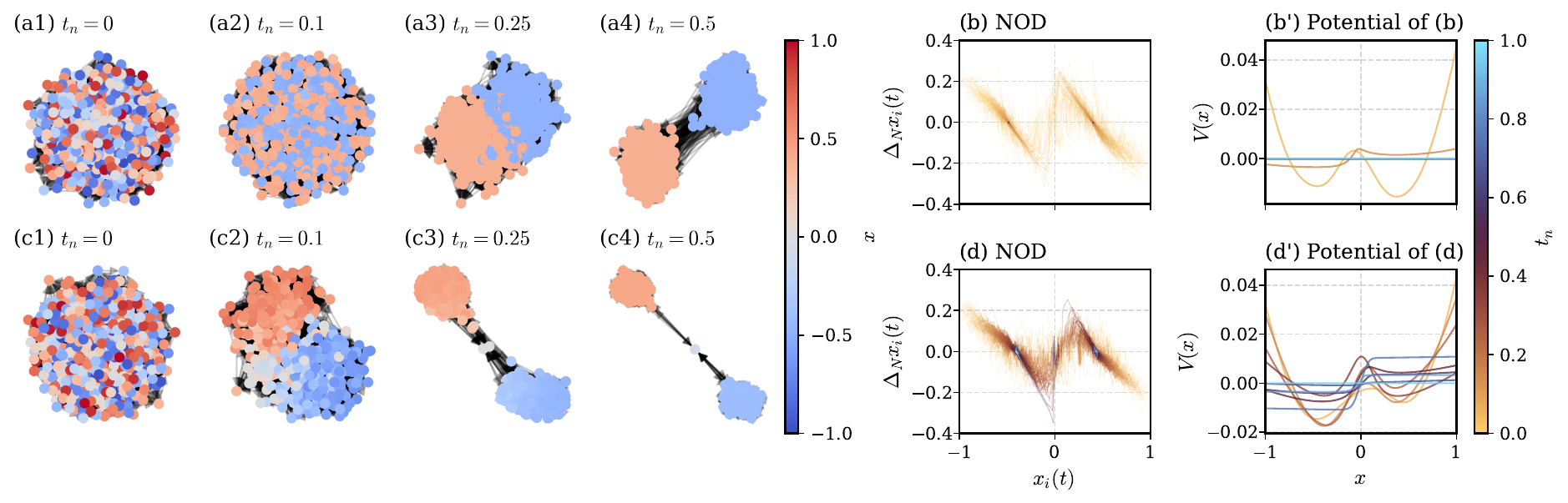}
    \includegraphics[width=0.95\textwidth]{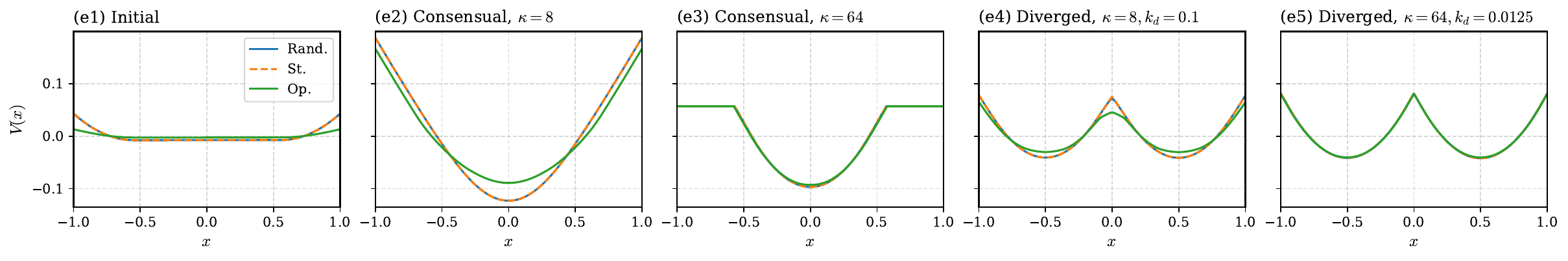}
    \caption{
    (a), (c) Evolution of a random network (500 agents) under random recommendation. (a) PbS pathway ($\alpha = 0.05, q=0.025$); (c) SbP pathway ($\alpha = 0.005, q=0.025$).
    (b), (d) Microscopic dynamics showing Normalized Opinion Difference (NOD) vs. current opinion.
    (b'), (d') Effective potential landscapes $V(x)$ derived from (b) and (d). Color gradient indicates normalized time $t_n$, showing the formation and flattening of the double-well potential.
    (e1)---(e5) Theoretically computed potential landscapes for three idealized network states,
under Random (Rand.), Structure-based (St.), and Opinion-based (Op.) recommendation. The standard deviations are determined as $\sigma_c = \kappa^{-1};\ \sigma_d = (2\kappa)^{-1}$.
     }
    \label{fig:toy-network}
\end{figure*}

In order to estimate the mean social force, we apply Gaussian kernel smoothing ($h=0.1$) to individual forces, i.e. $\bar{F}(x)=\frac{\sum_j K_h(x-x_j)F_j}{\sum_j K_h(x-x_j)}$, and integrate to obtain the time-varying potential. Phase portraits $(x_i,\Delta_n x_i)$ in Figs.~\ref{fig:toy-network}(b),(d) show agents relaxing toward attractors like damped particles, confirming the theoretical prediction. The corresponding time-varying potentials (Figs.~\ref{fig:toy-network}(b'),(d'), color-coded by normalized time $t_n=t/t_a$) reveal the formation and collapse of potential wells over time.

\paragraph{Mechanism.} The divergence between PbS and SbP arises from competition between the \emph{agent velocity} (governed by $\alpha$) and the \emph{collapse rate of the potential field}:
\begin{itemize}
    \item \textbf{High $\alpha$ (PbS, Fig.~(a)):} Responsive agents rush into potential wells before significant rewiring occurs; potential forms and collapses rapidly (swift flattening in Fig.~(b')), releasing social tension through fast polarization.
    \item \textbf{Low $\alpha$ (SbP, Fig.~(c)):} High-inertia agents move slowly, prolonging social tension and allowing structural rewiring (social selection) to produce segregation before opinions fully polarize.
\end{itemize}

In summary, whether a society polarizes or segregates first is determined by the interplay between individual inertia and the dynamic evolution of the social force potential.

\paragraph{Discussion} In theoretical calculations, the height of the potential well generated by the system is the same for different values of $\sigma$'s; however, in simulation results, the heights of the potential barriers and wells vary depending on the time. This is rational, because in theory we are essentially assuming that if there is a point at a certain location, the magnitude of the social force at that point should be this value; however, in numerical calculations, if there is no point at a certain location, the estimate of the social force will be closer to 0, leading to a gradually collapsing potential well.

\subsection{The Impact of Recommendation Algorithms on Evolutionary Pathways}\label{sec:ec-pathways}

\begin{figure*}[!htbp]
    \centering    
    (a)\raisebox{-\height}{\includegraphics[width=0.42\textwidth]{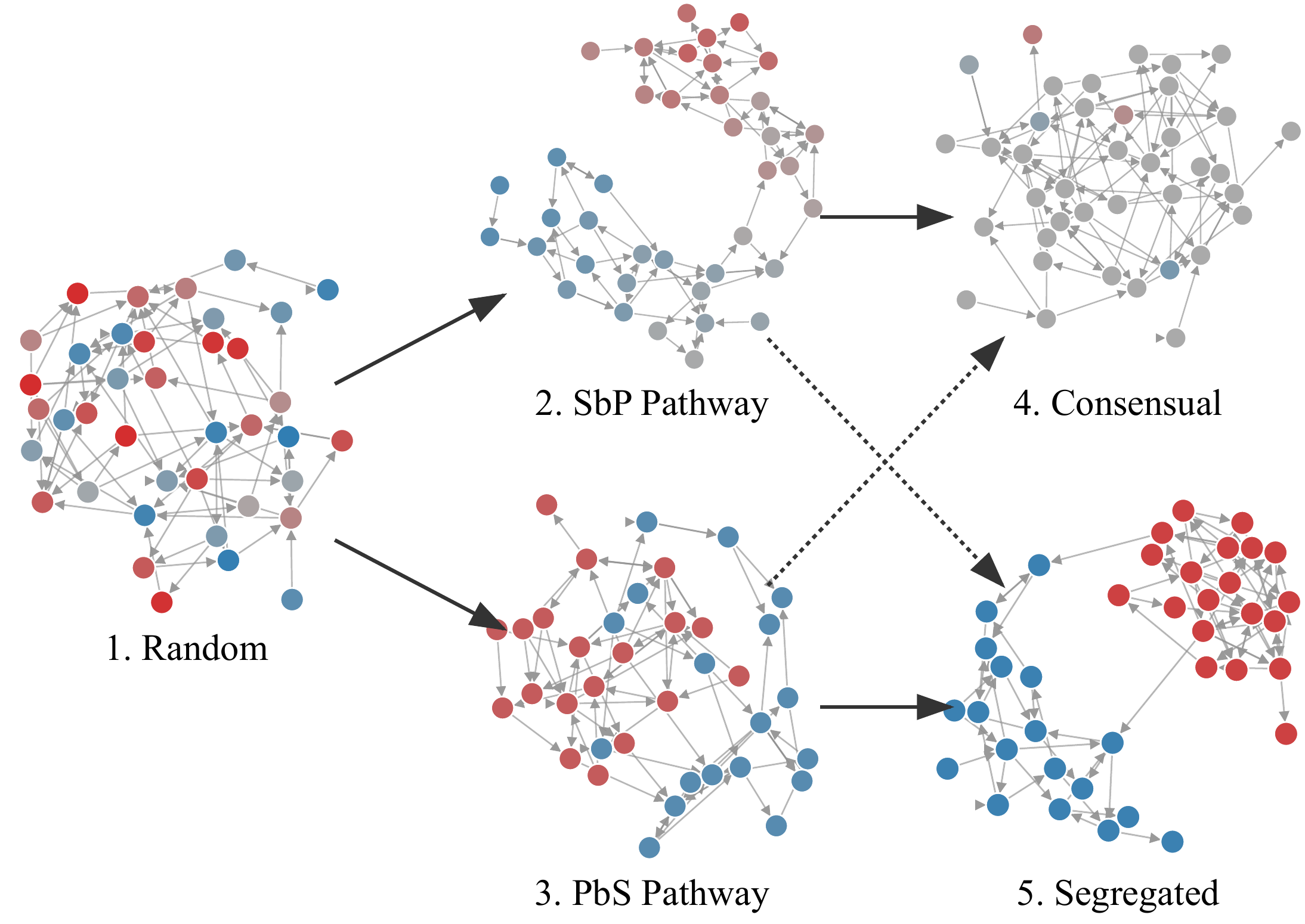}}
    (b)\raisebox{-\height}{\includegraphics[width=0.42\textwidth]{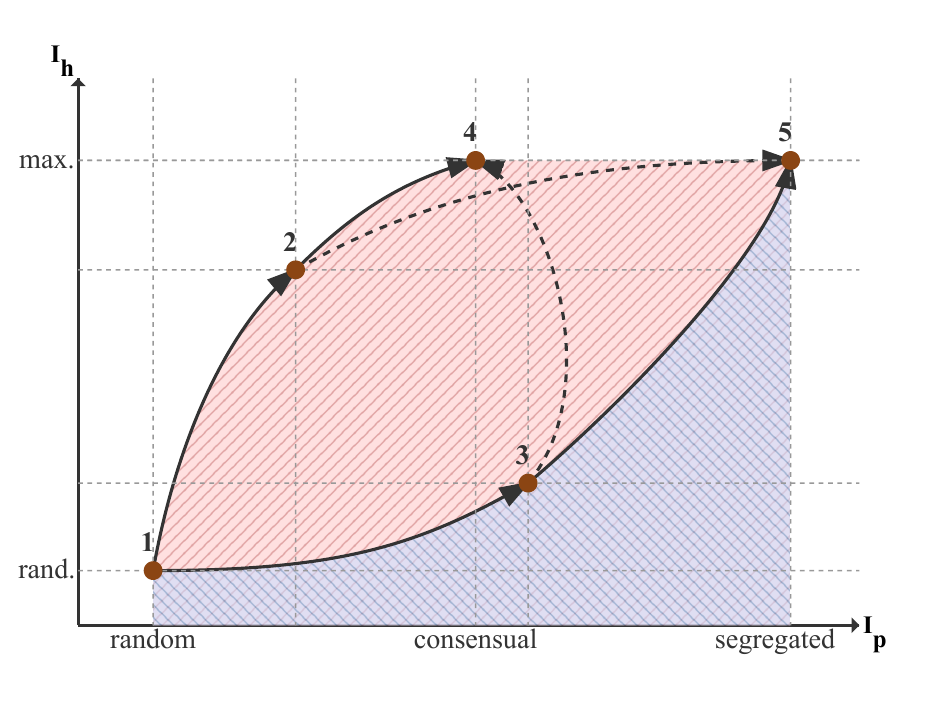}}
    \caption{
     (a) Typical evolutionary pathways of the society model. 
     From a (1) random state, the model evolves to (4) consensus or (5) segregation via (2) SbP or (3) PbS pathways.
     (b) Trajectories corresponding to these different pathways plotted on the $I_p$-$I_h$ diagram.
     }
    \label{fig:ec-pathways-illust}
\end{figure*}

\begin{figure*}[!htbp]
    \centering
    \includegraphics[width=0.8\textwidth]{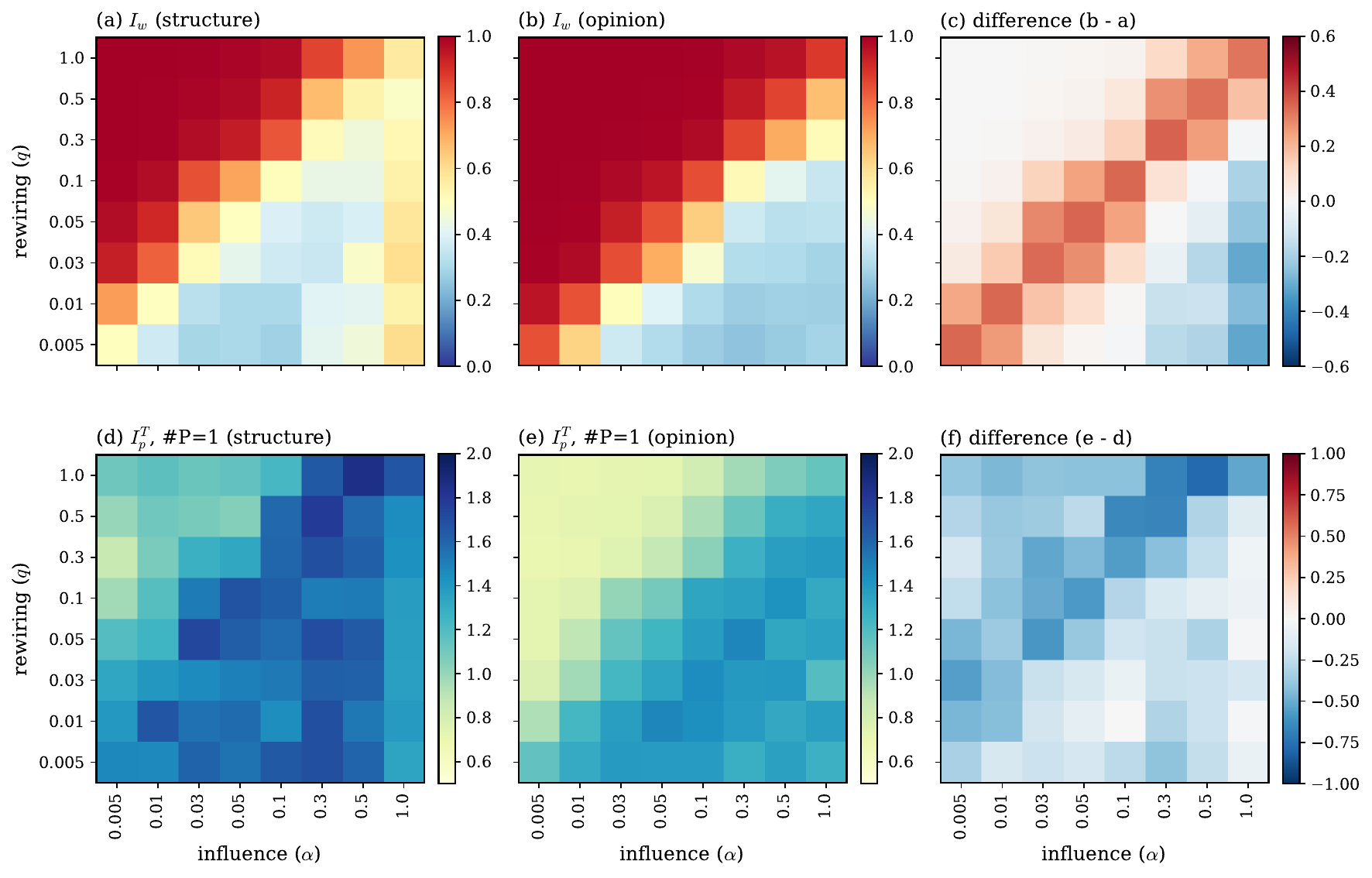}
    \caption{
    Heatmaps of simulation results.
    (a--c) Pathway index $I_w$.
    (d--f) Polarization trajectory length $I^T_p(t_a)$ for consensual scenarios ($\bar{I_p} \approx 0.062$ for structure-based, and $0.061$ for opinion-based cases). 
    Columns: (a, d) structure-based, (b, e) opinion-based, (c, f) differences.
    $I^T_p(t_a)$ are omitted for segregated cases as well as $I^T_h(t_a)$, as they consistently remain near 1.
    }
    \label{fig:cont-heatmap}
\end{figure*}

While the model ultimately converges to either consensus or polarization, the transition dynamics depend heavily on the balance between homophily and polarization (Figure \ref{fig:toy-network}). This trade-off drives the network along divergent pathways toward extreme states, as illustrated in Figure \ref{fig:ec-pathways-illust}. To analyze these dynamics, we simulated 1024 scenarios (36 trials each, fixed $\epsilon = 0.45$) comparing opinion-based versus structure-based recommendations. We utilize the \emph{trajectory length}, defined as $f^T(t) := \| \dot{f} \|_{L^1([0,t])}$, to quantify oscillation. Note that for segregated cases, trajectory indices remain near unity and are therefore omitted.

We categorize the parameter space into five regimes based on $D := \log_{10} q - \log_{10} \alpha$, ranging from \emph{rewiring-paramount} ($D \approx 2$) to \emph{influence-paramount} ($D \approx -2$). As extreme regimes are rare in real-world networks, we focus on the intermediate cases. 
In balanced scenarios ($D \approx 0$), opinion-based recommendations bias evolution toward SbP pathways (quantified by $I_w$, Figure \ref{fig:cont-heatmap}). Conversely, in certain influence-paramount cases, they drive evolution toward PbS pathways. Microscopic analysis suggests this counterintuitive result arises from the abrupt merging of two attractors formed via indirect information transmission (see Supplementary Materials). Furthermore, the stability of $I_w$ across varying repost rates ($p$) confirms its robustness for estimating the $\alpha$ and $q$ proportions.

The polarization trajectory length, $I^T_p(t_a)$, reveals that most consensual scenarios undergo transient polarization, with values reaching as high as 2. As shown in Figure \ref{fig:cont-heatmap}(f), $I^T_p(t_a)$ highlights significant algorithmic differences in balanced and rewiring-dominant regimes. Lower index values indicate convergence to consensus without traversing high-polarization regions, suggesting that opinion-based recommendations effectively mitigate transient polarization.

Synthesizing this evidence, we draw the following conclusions:
In societies where the tendencies of social influence and rewiring are comparable, opinion-based recommendation systems drive evolution along SbP pathways. 
Furthermore, for societies that ultimately reach consensus, opinion-based recommendations tend to steer pathways away from high-polarization zones.
This serves as the answer to (\textbf{RQ1}).

\subsection{Societal Interpretation and Implication of Pathway Index}\label{sec:ec-societal-interp}

\begin{figure*}[!htbp]
    \centering
    \includegraphics[width=0.9\textwidth]{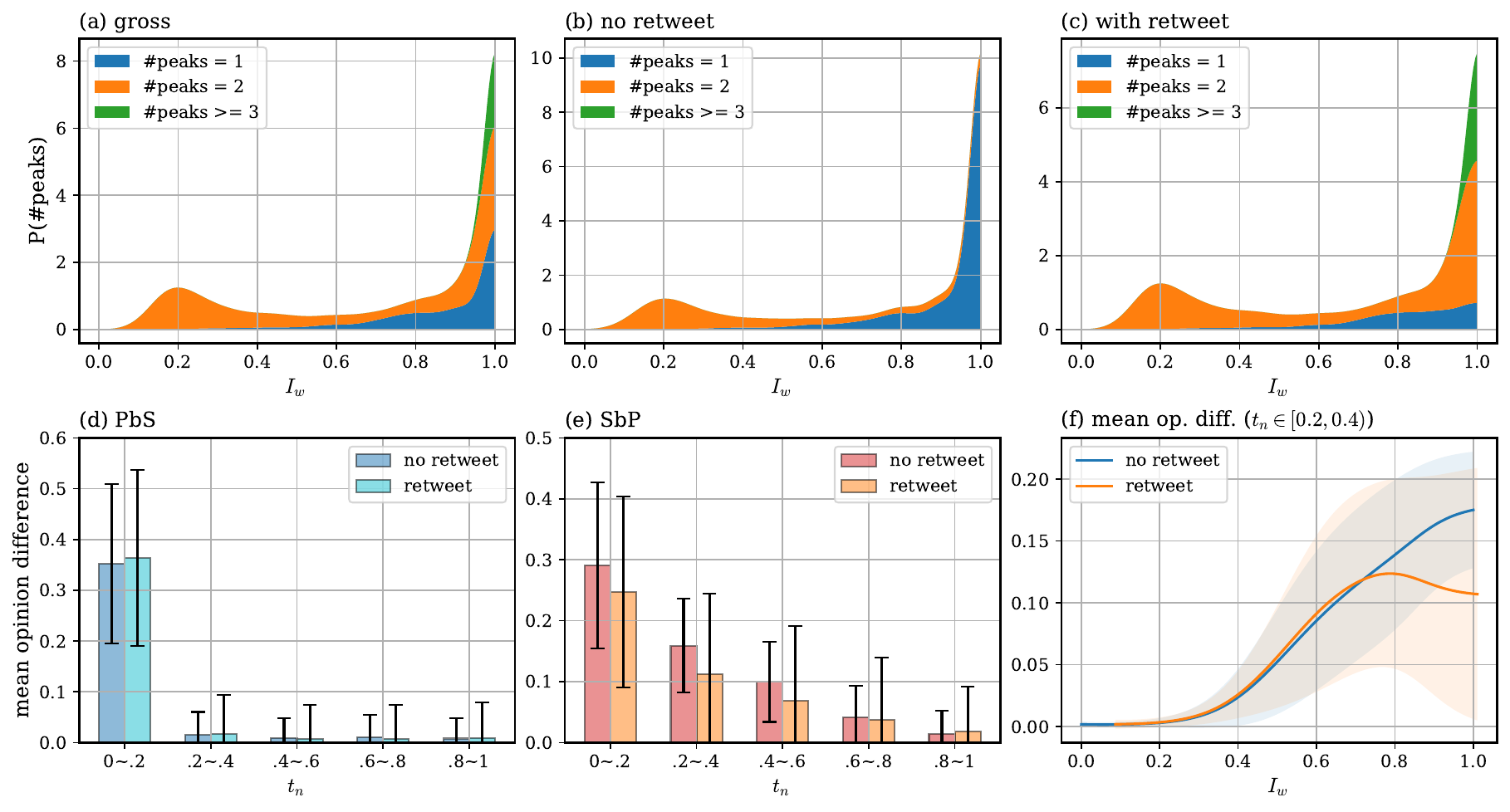}
    \caption{Impact of Reposting on Evolutionary Outcomes.
    (a--c) Probability Density Functions (PDF) of the pathway index $I_w$, estimated via Gaussian KDE. Colors indicate the final state: \emph{Consensual} (1 peak) or \emph{Polarized} ($\ge 2$ peaks).
    (a) shows the aggregate bimodal distribution, while (b) and (c) compare scenarios without reposting ($p=0$) and with reposting ($p>0$). 
    (d, e) Temporal evolution of Social Force for (d) PbS and (e) SbP cases.
    (f) Aggregated view for SbP cases, showing that reposting lowers the mean opinion shift but drastically increases the variance.}
    \label{fig:cluster-and-repost}
\end{figure*}

To answer \textbf{(RQ2)} and \textbf{(RQ3)}, we examine information sharing (reposting) using $I_w$ as a classifier. 
Figure \ref{fig:cluster-and-repost}(a) confirms $I_w$ is bimodal with a valley at $I_w \approx 0.6$, justifying a binary classification: PbS ($I_w < 0.6$) and SbP ($I_w \ge 0.6$).

\subsubsection{The Destabilizing Effect of Reposting in SbP Pathways}

Comparing Figures \ref{fig:cluster-and-repost}(b) and (c) reveals a critical divergence in outcomes. In the absence of reposting ($p=0$), SbP pathways (high $I_w$) almost invariably lead to consensus (indicated by the dominance of the blue area). However, the introduction of reposting ($p > 0$) fundamentally alters this dynamic, resulting in a significant proportion of polarized outcomes (orange/green areas) within the SbP regime.

To understand the microscopic mechanism driving this shift, we analyze the \emph{Mean Opinion Difference}—which serves as a proxy for the magnitude of the Social Force—and its variance over normalized time $t_n$.
As illustrated in Figures \ref{fig:cluster-and-repost}(d) and (e), the impact of reposting is non-uniform. In PbS cases (d), the dynamics remain largely unaffected. Conversely, in SbP cases (e), reposting induces a \emph{high-variance, low-mean} dynamic. While the average opinion change decreases (suggesting agents are locally trapped), the standard deviation explodes compared to the no-repost baseline.

\paragraph{Discussion}
From a potential energy perspective, these high fluctuations (large standard deviation) imply that reposting \emph{roughens} the energy landscape.
In the early stages, reposting acts as a bridge, introducing long-range interactions that prevent the potential field from relaxing into a single global well. Instead of a smooth collapse toward consensus, the information noise creates a complex landscape with multiple local minima separated by high potential barriers.
In SbP scenarios—where agents already possess high inertia (low susceptibility)—these local barriers are sufficient to trap clusters of agents.
Consequently, rather than facilitating global convergence, reposting solidifies local homogeneity, effectively turning what would have been a consensual society into one fragmented by entrenched echo chambers.

\subsubsection{The Effect of Pathways on Echo Chambers}

\begin{figure*}[!htbp]
    \centering
    \includegraphics[width=0.75\textwidth]{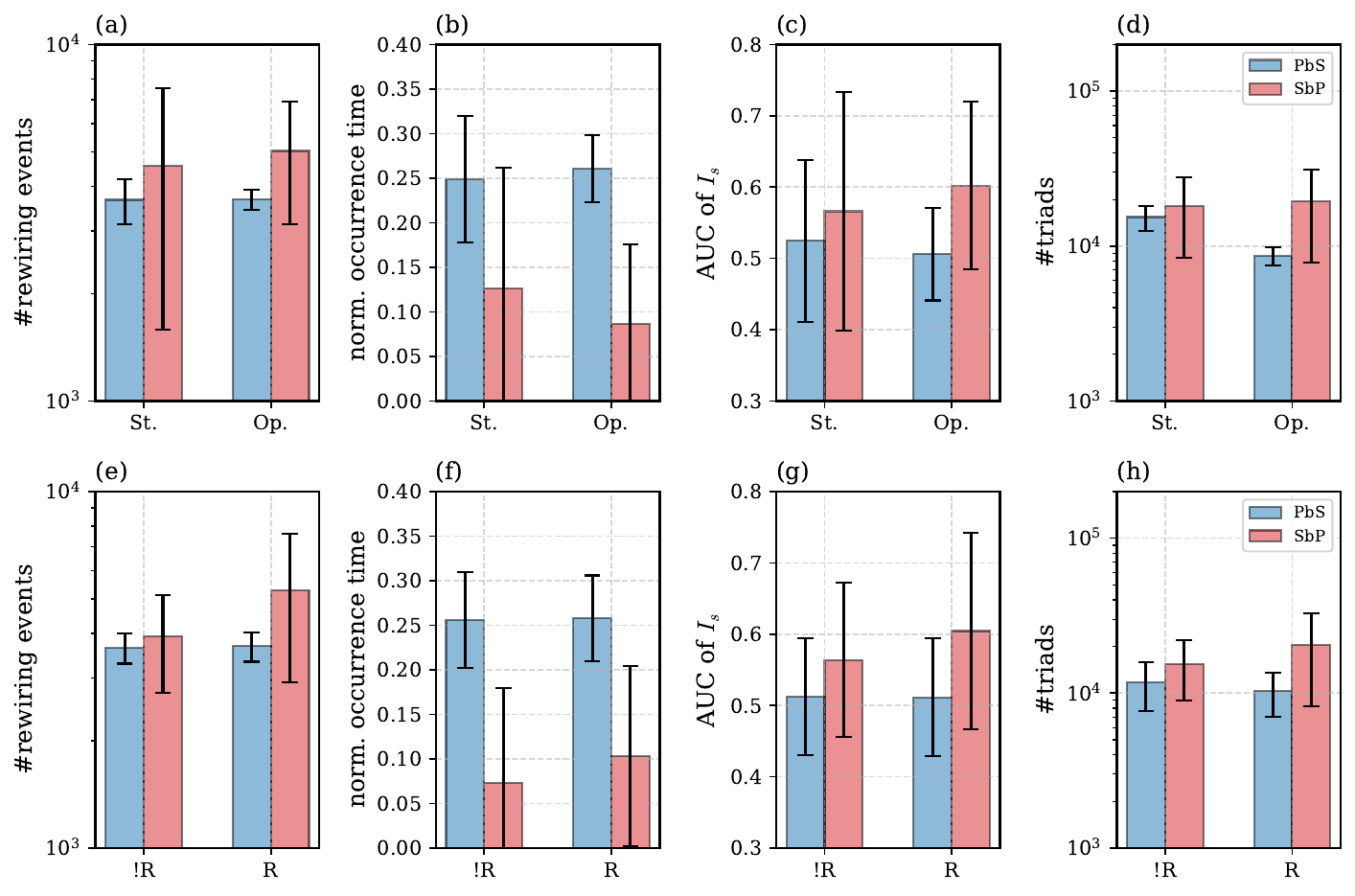}
    \caption{
    Comparison of feature indices across different algorithmic and societal characteristics.
    Panels show the relationship between simulation conditions and:
    (a, e) the total count of rewiring events;
    (b, f) the normalized occurrence time of rewiring events;
    (c, g) the AUC of the subjective polarization index $I_s$; and
    (d, h) the final number of closed triads.
    Top row (a--d) aggregates results by recommendation algorithms (Structure- vs. Opinion-based), while bottom row (e--h) aggregates by reposting behavior (!R: no repost; R: repost). Error bars indicate standard deviations.
    }
    \label{fig:cont-interpret}
\end{figure*}

Beyond the distribution of $I_w$ and opinion peaks, the properties of echo chambers at both individual and collective levels are also critical. We use rewiring events as a proxy to quantify the tendency of a society to insulate itself from discordant opinions at the individual level, measured by their frequency and normalized time of occurrence. As seen in Figure \ref{fig:cont-interpret}(a, b, e, f), the event count and average time remain relatively stable in PbS cases ($I_w < 0.6$). In SbP cases ($I_w \geq 0.6$), the situation is more complex: the presence of reposting strongly increases the number of events (+34\%). Furthermore, an opinion-based recommendation algorithm decreases the normalized occurrence time by 32\%, whereas allowing reposts increases it by 41\%. In the later stages of model evolution, as the society becomes highly homogeneous, rewiring events become rare. However, the introduction of reposts can create scenarios where a reposted post conflicts with an agent's opinion, even if the original author is a concordant followee, leading to later and more frequent rewiring events as observed in Figure \ref{fig:cont-interpret}(g). This reveals a potential risk of opinion-based recommendation systems: they can algorithmically entrench individual-level homophily.

At the collective level, we use the prevalence of closed triads (topological structure A→B, A→C, and B→C), considered a minimal unit of an echo chamber \cite{jasny2015empirical}, to assess the collective state. 
We found that both the use of opinion-based algorithms and the encouragement of retweeting behavior cause path tendencies to be associated with the final closed triangle structures. As shown in Figure \ref{fig:cont-interpret}(d), for the group using the structure-based recommendation algorithm, the difference in the number of closed triangles between the two paths is only 17\%, whereas for the opinion-based recommendation algorithm, this difference reaches 123\%. 
As shown in Figure \ref{fig:cont-interpret}(h), if retweeting is prohibited, the difference is 32\%, but when retweeting is allowed, it reaches 98\%.
Furthermore, it is particularly worth noting that, in terms of absolute numbers, the number of closed triangles in communities on the SbP path using opinion-based recommendations is 7\% higher than those using structure-based recommendations. This indicates that, under these conditions, the opinion-based recommendation algorithms are already capable of exacerbating the severity of the collective-level echo chambers to a degree slightly exceeding that of the structure-based ones---despite the fact that the structure-based algorithms are explicitly designed to maximize echo chambers, whereas the opinion-based ones are not.

While the intensity of echo chambers is a key outcome, their formation speed is an equally important counterpart. Figure \ref{fig:cont-interpret}(c, g) illustrates the relationship between $I_w$ and the AUC of subjective polarization index $I_s$ among $t \ [0, t_a]$.
A large AUC value indicates that the clustering process is faster than the overall average, implying a front-loaded evolution. 
This implies that for SbP cases, the network structure falls into an echo chamber more quickly, then gradually develops into an extremely potent echo chamber.

\ 

In summary, the answer to (\textbf{RQ2}) is as follows:
Information sharing selectively acts as a critical destabilizer in SbP pathways. By injecting high-variance information noise, reposting disrupts the system's natural relaxation toward global consensus, trapping users in local potential wells and amplifying the formation of entrenched echo chambers.

The answer to (\textbf{RQ3}) is as follows:
SbP pathways represent a \emph{front-loaded} societal fragmentation. They accelerate early individual isolation through frequent rewiring and rapidly compound into intense collective echo chambers (as quantified by closed triads)---an effect severely exacerbated by content-based algorithms and information sharing.

\section{Discussion}\label{sec:discussion}

The key feature of this model is the inclusion of a content recommendation algorithm by extending the dynamics of the Hegselmann-Krause (HK) model \cite{hegselmann2002opinion}---a typical bounded confidence model.
Our work also builds upon a rich body of research that explores variant social network models, focusing on features like network rewiring \cite{sasahara2021social} or recommendation algorithm \cite{santos2021link}.
By doing so, we mathematically articulate how algorithmic curation of information directly participates in opinion dynamics. We demonstrate that opinion-based recommendation systems, by their very design, generate homophily at an algorithmic level, which carries a significant potential to create and amplify echo chambers.

Notably, with the absence of preferential attachment of either recommendation or following behavior, the simulated model exhibit all the behaviors without showing the emergence of power-law distributions, which are characteristic of many real-world social networks.
This aligns with \citeauthor{galam2024spontaneous}'s work demonstrating that echo chambers emerge naturally at the end of a local updating process among agents who initially hold different opinions with the absence of preferential attachment\cite{galam2024spontaneous}.

A particularly interesting and unforeseen finding is the dramatic impact of the repost mechanism (controlled by parameter $p$) on the formation of echo chambers. The presence or absence of reposting leads to substantial differences across various sides of echo chambers. The drastic difference of opinion peak's count yielded by solely allowing or disallowing reposting in rewiring-paramount cases, as shown in Figure \ref{fig:cluster-and-repost} (c), coincides \citeauthor{conover2011political}'s conclusion that retweeting leads to homogeneous groups while debates and mentions may help maintain cross-group engagement \cite{conover2011political}.
While some existing models have incorporated reposting behaviors \cite{sasahara2021social, yan2021understanding, larooij2025fix}, a comprehensive comparative study has been lacking. This research partially fulfills that gap by treating the repost probability as a core model parameter, revealing its critical role in the interplay between influence and selection.

In addition, our finding that reposting acts as a critical destabilizer is powerfully corroborated by concurrent research utilizing fundamentally different modeling architectures\cite{oliveira2026mechanistic}, demonstrating that a low innovation probability (analogous to high reposting rate $p$) drastically reduces content diversity and locks the network into polarized states (analogous to a collapsed potential landscape).
The convergence of these independent methodological approaches provides robust evidence that circulating existing content, rather than generating new ideas, is a primary driver of echo chamber entrenchment.

\subsection{Difficulty and Possibility of Eliminating Echo Chambers}

The proposed model, in its current form, offers some rather pessimistic conclusions for dismantling echo chambers that are already deeply entrenched, analytically and numerically.
The experimental results support the analogy drawn in Section \ref{sssec:theoretical} between the simulation process and non-equilibrium relaxation following quenching.
The model evolves from a critical state, immediately collapsing into several potential wells.
At the same time, when the influence parameter $\alpha$ is relatively strong, the model can reliably end in consensus along the PbS pathway. This can be analogized to domain wall absorption.
These mechanisms makes the ability to escape the wells diminish over time.
Many empirical and modeling studies corroborate the finding that once opinion clusters become highly polarized and segregated, they are incredibly resilient to change \cite{piccoli2018sparse, hata2025manipulating}. 

On the other hand, there are studies suggest that the emergence and subsidence of social issues follow a cyclical pattern\cite{lorenzspreen2019accelerating}, or that there are fluctuations in individuals’ information sources\cite{fletcher2018are, fletcher2018automated}. This suggests that the society may undergo repeated quenching, and that even after convergence, a certain background temperature persists, keeping the individuals minimally active.

Therefore, the goal should not be the complete elimination of echo chambers---which can, to some extent, be a natural result of human nature and association---but rather to explore mechanisms that can control their intensity and prevent them from reaching pathological levels. 
This raises two further questions: 1. How can we ensure that the society deviates as much as possible from the \emph{worst-case} (like a state of extremely low free energy)? 2. Does the society exhibit a disordered state above the critical temperature, and does it possess background temperature?
This necessitates further empirical research to understand which human psychological processes might be leveraged to dive deeper in the dynamics of echo chambers.

\subsection{Implications for Social Mitigation and Intervention}

In the context of a social media platform, the recommendation algorithm acts as a controllable input, while the desired goal is to maintain network health---for instance, by minimizing polarization or maximizing the diversity of opinions to which users are exposed. This model provides a framework for designing such interventions.

The model is particularly well-suited for modeling and monitoring the evolution of discussions around topics with a dualistic nature, whether the national partisan divide or minor events within specific communities.
By analyzing the network structure and estimating user opinions, platform operators could use the indices proposed in this paper---the subjective polarization index ($I_s$) and the pathway index ($I_w$)---to assess the state of the discourse and decide when and how to intervene. The subjective polarization index $I_s(t)$ tracks the real-time \emph{activity level} of opinion change, while the pathway index $I_w$ estimates the network's underlying tendency to favor either social influence or social selection (rewiring).

The intervention opportunities and mitigation strategies lie in the stage-dependent application of different recommendation strategies, guided by these indices:

\paragraph{Early-Stage Intervention for Consensus.}
In the initial phase of an event's discussion, if the goal is to foster consensus or prevent premature fragmentation, an \textbf{content-based recommendation system} can be effective. By recommending content that is diverse yet still within users' confidence bounds, it can gently guide opinions toward a central point. However, this should be paired with a mechanism that \textbf{discourages reposting} (i.e., reduces the visibility of reposts in recommended feeds). Our results show that high repost rates ($p$) in conjunction with content-based recommendations can rapidly accelerate polarization.

\paragraph{Managing Polarization in Rewiring-Prone Networks.}
The network goes through a phase in which the walls of the echo chamber gradually take shape; this is the last effective window of opportunity to shape public opinion.
If it undergoes PbS pathway, \textbf{link-based recommendation system} is effective for pulling back the opinion from polarization.
If it exhibits strong SbP tendency, combining an \textbf{content-based recommendation system with the encouragement of reposts} can paradoxically be beneficial by guiding the network toward a state of multi-polarization rather than bipolarization. 
Reposting in this stage serves to bridge or divide nascent clusters by different cases.

\paragraph{Maintaining Network Vitality and Preventing Ossification.}
In later stages, or for networks that are becoming static, the main concern may be the ossification of the network structure, characterized by a high count of closed triads and a low subjective polarization index ($I_s$). To counteract this and maintain a healthy flow of information, deploying a \textbf{link-based recommendation system} is advantageous. By suggesting new connections based on shared neighbors rather than opinion similarity, this strategy can introduce structural diversity and break down the walls of established echo chambers, thereby increasing the potential for cross-cutting exposure over the long term---although its effect over echo chambers remains controversial\cite{bail2018exposure, garimella2018political}.

\ 

These intervention strategies are subtle; they \emph{go with the flow} of user behavior rather than imposing drastic changes like content removal or censorship. This subtlety is a significant advantage for practical implementation, as it minimizes disruption to the user experience. However, the effective use of these indices requires empirical calibration. Their absolute values are less meaningful than their comparative values across different events on the same platform.

\subsection{Possible Future Works}

Building on this research, we propose three primary directions for future investigation:

\paragraph{Empirical Validation.} 
To verify practical utility, a critical next step is to conduct empirical studies to validate the model and its proposed intervention strategies. 
Calibrating agent-based network models against empirical cascade distributions from major polarizing events offers a viable method to map theoretical parameters (like $p$ and $\alpha$) to observable platform dynamics\cite{oliveira2026mechanistic}.
It is essential to test the effectiveness of the $I_s$ and $I_w$ indices in characterizing the evolution of online discussions under calibrated parameters for bridging the gap between theoretical modeling and practical application.

\paragraph{Enhancing Model Granularity and Depth.} 
Real-world opinion formation is far more complex than the one-dimensional bounded confidence model suggests. Future research should aim to increase model fidelity by incorporating multi-dimensional opinion vectors, heterogeneous agents\cite{friedkin1990social, friedkin1999social, galam2023unanimity}, partisan sorting effects \cite{tornberg2022digital}, and meta-opinions \cite{yan2024incorporation}. The integration of Large Language Models (LLMs) as agents offers a promising path to simulate complex cognitive behaviors while maintaining model simplicity \cite{gao2023s3, ziems2024large, larooij2025fix}. 
Furthermore, deeper investigation is needed into the system's dynamic characteristics. The current model exhibits underdamped behaviors leading to \emph{oscillations}, which resemble public opinion cycles or social movements \cite{pluchino2004changing, xue2020opinion}. Future work should simulate realistic recommendation algorithms that include feedback slack and delays, as these imperfections create ``social forces'' that significantly impact opinion mitigation\cite{xiao2019beyond, hassan2019trust, wu2022survey, ren2014random}.

\paragraph{Engineering and Metric Refinement.} 
To apply this framework to actual social governance, the analysis indices require engineering refinements. The current indices, while analytically useful, exhibit limitations in broader conditions. For instance, the polarization index $I_p$ loses descriptive power in multi-polarization scenarios (beyond bipolarity), and the activity time $t_a$ and pathway index $I_w$ are currently computable only after the society reaches a static state. 
Future work should focus on developing generalized indices that can be computed in real-time based on potential energy and estimated social forces. This would enable dynamic monitoring and timely interventions before echo chambers become deeply entrenched.

\section*{Acknowledgments}

\paragraph{Author Contributions}

J. Z. and K. S. designed the study. J. Z. performed the modeling, simulation and result analysis. J. Z. (mainly), K. S. and Y. C. wrote the manuscript. All authors reviewed and approved the final manuscript.

\paragraph{Code Availability}

The code used to generate the results in this study is available at 
\url{https://github.com/billstark001/social-media-models} (simulator) and \url{https://github.com/billstark001/extended-hk-model} (data analysis). 
Additional details on the implementation and usage are provided in the repository's README file.

\paragraph{Declaration of Generative AI Usage}

\textit{Claude Sonnet} and \textit{Gemini Pro} series are applied \emph{strictly to translate and to polish some of the original Chinese and Japanese draft}. All generated text are manually proofread and corrected per-sentence by the authors.

    % Print references for the main text
    % \nocite{*}
    \printbibliography[heading=subbibliography, title={References}]

\end{refsection}

\clearpage

% =========================================
% SUPPLEMENTARY MATERIALS
% =========================================

\appendix

% --- Reset Counters for Supplement ---
\setcounter{page}{1}
\renewcommand{\thepage}{S\arabic{page}}
\renewcommand{\thefigure}{S\arabic{figure}}
\renewcommand{\thetable}{S\arabic{table}}
\setcounter{figure}{0}
\setcounter{table}{0}

% --- Supplementary Header ---
\begin{strip}
    \begin{center}
        \LARGE \textbf{Supplementary Materials}
    \end{center}
\end{strip}

% --- Supplementary Reference Section ---
\begin{refsection}

    % \documentclass[main.tex]{subfiles}
% \usepackage{xr-hyper}
% \externaldocument{main}

% \title{Supplementary Materials}
% \author{}

% \begin{document}

% \maketitle

Here, we provide additional figures, explanations, and experiment details
that support the main text.

\section*{Simulation Details}

\subsection*{Simulation Setup and Parameters}

All simulations were conducted on randomly generated directed graphs with $n=500$ agents. The graphs were generated using an Erd\H{o}s-R\'enyi random digraph model with a connection probability $p = k_o / (n-1)$, where the average out-degree was fixed at $k_o=15$. This fixed out-degree represents users' limited attention. Initial agent opinions were drawn from a uniform random distribution on $[-1, 1]$. A simulation is considered to have converged and is halted if there is negligible opinion change and no network structure change for 60 successive steps, or if it exceeds a maximum of 20,000 steps. 
% Is this necessary? (I suppose not...)
% Unless otherwise specified, simulations used parameters $\alpha=0.05, q=0.05, p=0.1$.

For different recommendations, opinion- and structure-based ones are used (random ones are excluded since it does not signify so much). For the opinion-based recommendation algorithm, the historical memory length $k_h$ is set to 0, 2, 6. The parameters related to different societies are chosen from different ranges, specifically $\epsilon = 0.45$, $\alpha$ and $q$ range from 0.005 to 1, $p$ range from 0 to 0.5. These yield $4 \times 4 \times 8 \times 8 = 1,024$ scenarios. Each scenario is simulated for 36 times to eliminate randomness.

\subsection*{Parameter Selection: Confidence Boundary $\epsilon$}

\begin{figure*}[!htbp]
    \centering
    \includegraphics[width=0.9\textwidth]{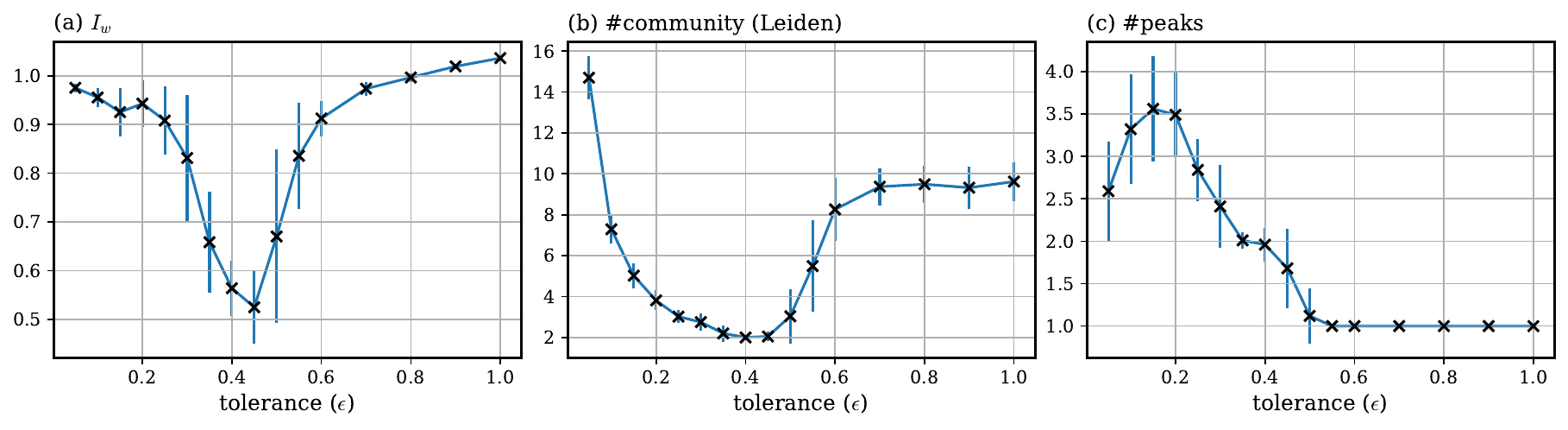}
    \caption{The final (a) pathway index $I_w$, (b) community count and (c) opinion peaks count across 100 simulations for $\epsilon = 0.05, 0.1, ..., 1$, with random recommendations and parameters $\alpha=0.05, q=0.05, p=0.1$. the community and opinion peaks count approaches 2 around $\epsilon = 0.4$, and around the same point the pathway index approaches its balanced value as well (albeit it is flawed for multi-polarization cases). Due to the need to including some consensual cases, $\epsilon = 0.45$ is chosen in the subsequent simulations.}
    \label{fig:supp:eps-select}
\end{figure*}

The confidence boundary $\epsilon$ is a critical parameter governing the scope of social influence. To select an appropriate value, we ran simulations for $\epsilon$ ranging from 0.05 to 1.0 (Figure \ref{fig:supp:eps-select}). The results show that key indicators of bipolarization, such as the number of opinion peaks and communities, consistently approach a value of 2 around $\epsilon \approx 0.4$. This finding aligns with previous work showing that bounded confidence models are highly susceptible to bipolarization around this threshold \cite{sasahara2021social}. To ensure our analysis captured dynamics in this critical region while also including some cases of consensus, we chose $\epsilon = 0.45$ for all subsequent simulations.

\subsection*{Reproducibility of Prior Work}

\begin{figure*}[!htbp]
    \centering
    \includegraphics[width=0.75\textwidth]{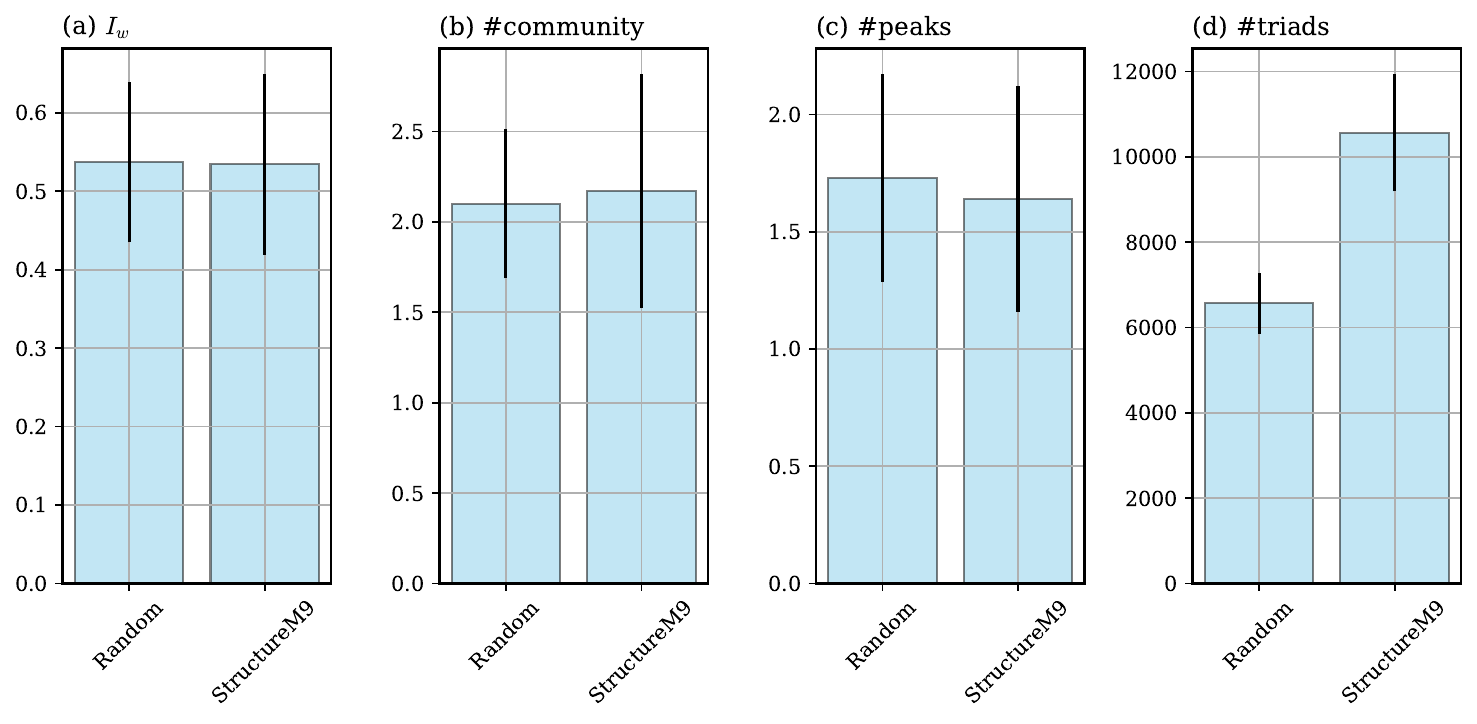}
    \caption{Some feature indices of the reproduction of \citeauthor{zignani2014link}'s result. Simulations are performed 100 times for a random and structure-based recommendation algorithm respectively, all with parameters $\alpha=0.05, q=0.05, p=0.1, \epsilon=0.45$. While other indices hardly differ, the average count of closed triads is 60\% larger than the random cases.}
    \label{fig:supp:reproduce-zignani}
\end{figure*}

To validate our model's fundamental dynamics, we tested its ability to reproduce established findings on the impact of network structure on social phenomena. Specifically, we examined the effect of a structure-based recommendation algorithm on the formation of closed triads, a result demonstrated by \citeauthor{zignani2014link}\cite{zignani2014link}. As shown in Figure \ref{fig:supp:reproduce-zignani}, our simulations confirm this effect. When applying a structure-based recommender (mimicking idealized link recommendations), the average count of closed triads was 60\% larger than in simulations with a random recommender, while other global indices remained comparable. This demonstrates our model's capacity to capture the structural reinforcement mechanisms that are foundational to echo chamber formation.

\section*{Additional Details of Results}

\subsection*{Convergence \& Activity Time Step}

In all scenarios, the model stops if there is almost no change in opinion (specifically, $\left|x_i(t+1) - x_i(t)\right| < 10^{-7}$ for all $i$'s) and strictly no change in network structure for successive steps $h$ (60 in this work), or it does not converge after too many steps (20000 in this work). Despite these conditions prevent models for performing infinite many useless steps, there are steps before the model converges where the society does not make any meaningful changes. 

To enable meaningful comparison, we define an \emph{activity time}, $t_a$, as the time point where the system becomes effectively static. 
Given the relationship of the subjective polarization index $I_s$ with information entropy, it can naturally be linked to the concept of \emph{temperature} when discussing the potential energy of a model. For example, we can define temperature as $T = 1 - I_s$, such that $T \approx 1$ when the system is in a random state and $T \approx 0$ when the model has essentially converged.
Therefore, one can define the activity time step $t_a$ as the time when the system almost chills. In the analyses, it is defined as the earliest time step that
\begin{equation*}
   I_s(t) \geq \max(0.98 * I_s(t_{\text{sim}}), 0.75)
\end{equation*}
is satisfied.
Here, $t_{\text{sim}}$ refers to the time step in which the simulation is stopped.
If the condition is never satisfied, $t_a$ is defined as $t_{\text{sim}}$.

This allows for the definition of a \emph{normalized time} $t_n: = \frac{t}{t_a}$, where $t_n = 0$ indicates the very beginning of simulation as $t$ does, and $t_n = 1$ indicates there is almost no meaningful changes in the simulated society.

\begin{figure*}[!htbp]
    \centering
    \includegraphics[width=0.96\textwidth]{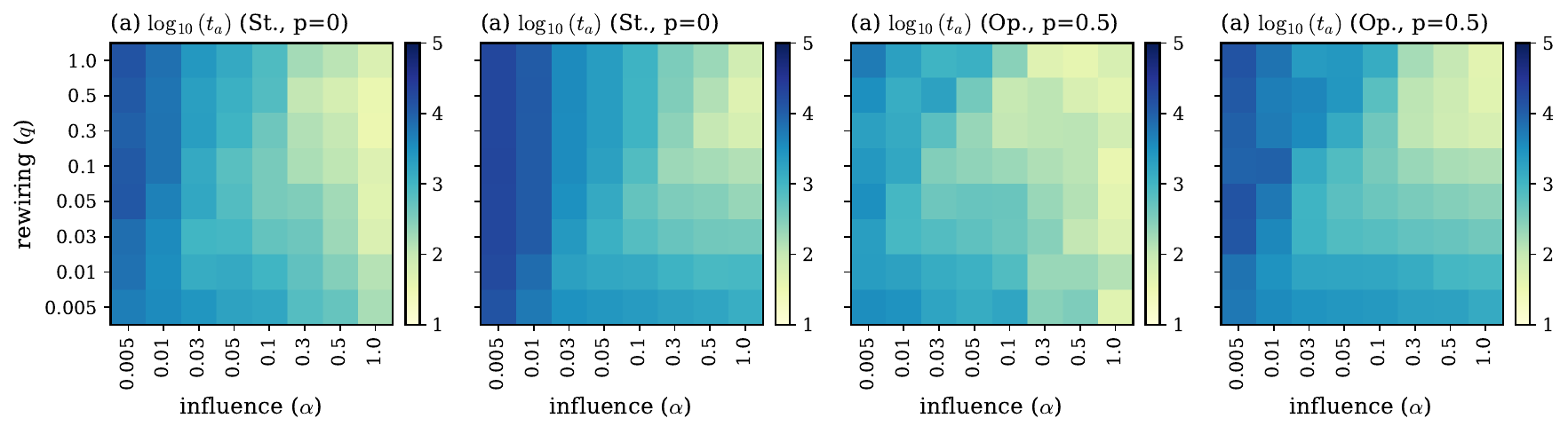}
    \caption{The heatmap of the logarithmic-scale activity time ($\log_{10}t_a$) across all simulation scenarios. Depending on the different parameters, $t_a$ varies from $10^{1}$ to $10^{5}$. 
    Overall, the $t_a$'s in opinion-based scenarios with $q > \alpha$ is larger than those with the same parameter settings in structure-based scenarios.
    }
    \label{fig:supp:active-time}
\end{figure*}

$t_a$ could vary from $10^1$ to $10^5$, as depicted in Figure \ref{fig:supp:active-time}. Larger repost rate sligntly accelerates the convergence for rewiring-dominant cases.

\subsection*{Pathways Along the $I_p$-$I_h$ Diagram}

\begin{figure*}[!htbp]
    \centering
    \includegraphics[width=0.75\textwidth]{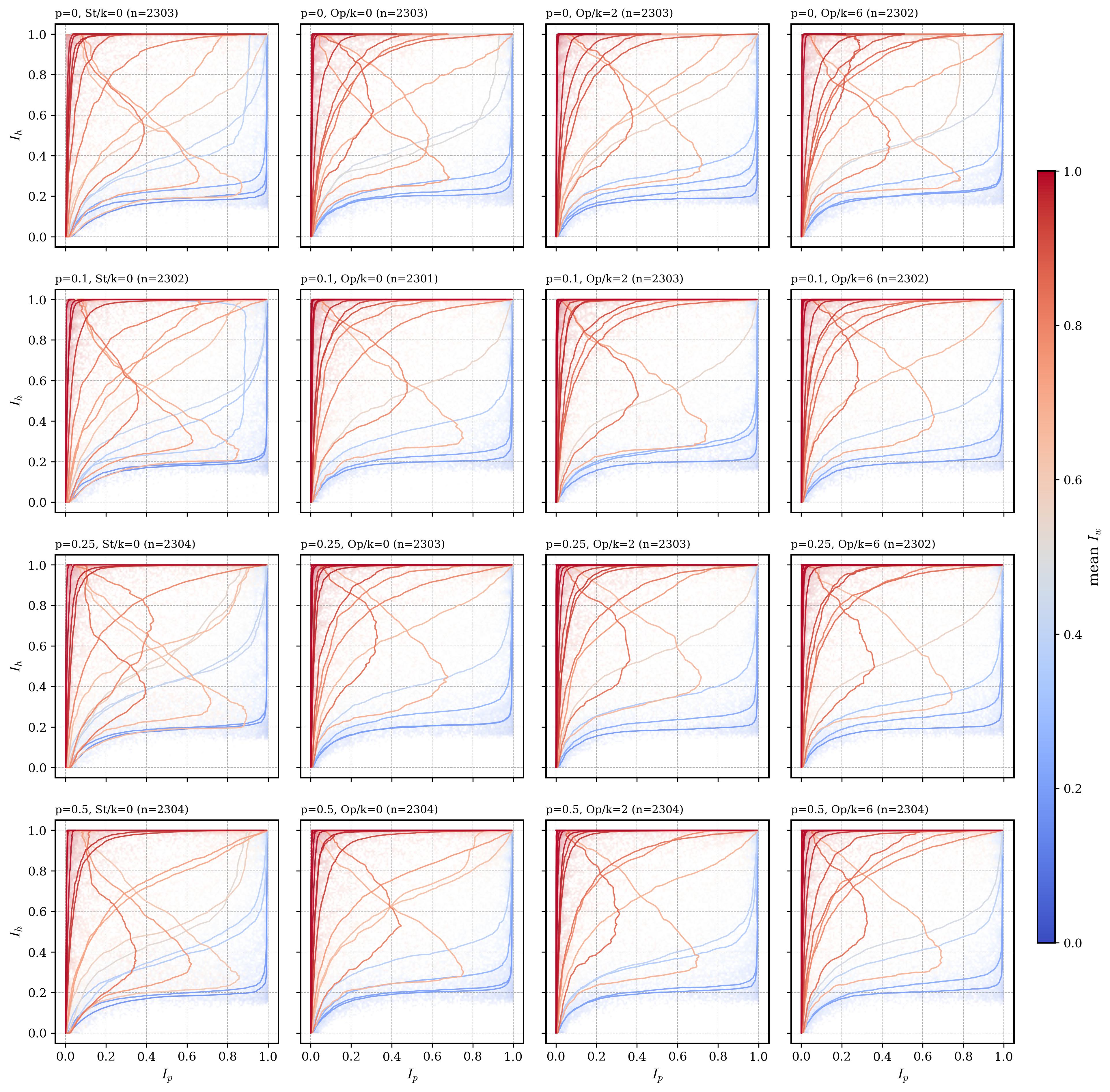}
    \caption{Trajectories of $I_p$ versus $I_h$ across various retweet rates and recommendation strategies. The underlying heatmaps visualize the log-scaled density distribution of all simulated trajectories, shaded by the local mean $I_w$. To illustrate the primary evolutionary paths, the solid lines show the median spatial trajectories of the dominant clusters within the data, with the line colors mapping to the average $I_w$ of each respective cluster.}
    \label{fig:supp:ip-ih-trajectory}
\end{figure*}

Figure \ref{fig:supp:ip-ih-trajectory} clearly shows several distinct typical community evolution trajectories and their densities, as well as the corresponding $I_w$ values for these typical trajectories, as illustrated in the main text (Figure \ref{fig:ec-pathways-illust}). If we focus on the trajectory that moves from the bottom-left corner to the right side of the space and then to the top-left corner, we can see that, for any forwarding probability, the maximum $I_p$ of the community using the structure-based recommendation algorithm is larger, and the corresponding $I_h$ is slightly smaller. This corresponds to the conclusions in Figure \ref{fig:cont-heatmap}(d)–(f) in the main text. Focusing on trajectories that do not pass through the right side of the space, we observe that allowing reposts increases the probability that SbP trajectories end in the top-right corner (i.e., opinion fragmentation). This corresponds to the conclusion in Figure \ref{fig:cluster-and-repost} in the main text: allowing reposts exacerbates the fragmentation of SbP trajectory communities.

\subsection*{Factors Affecting SbP Societies}

\begin{figure*}[!htbp]
    \centering
    \includegraphics[width=0.9\linewidth]{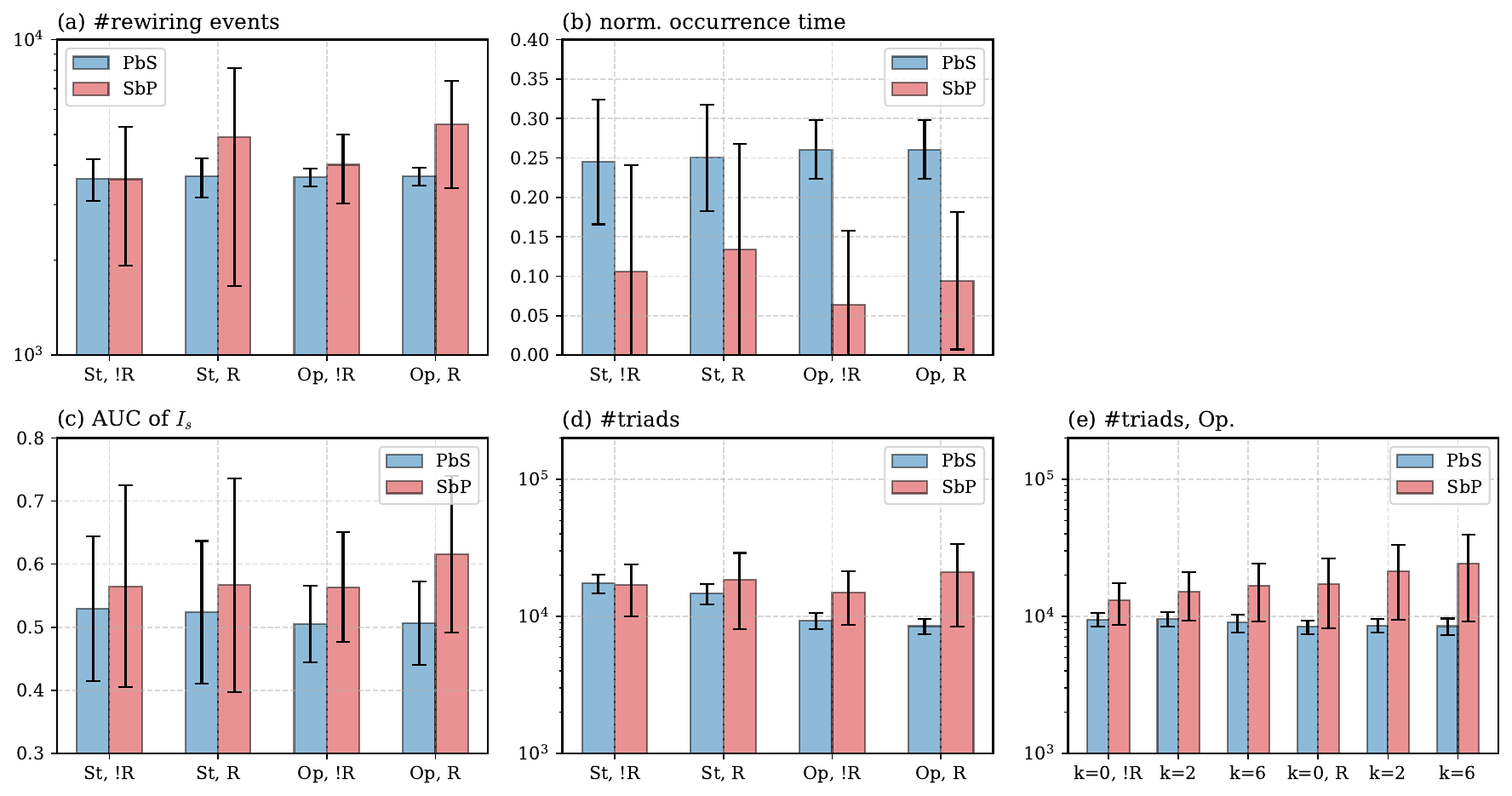}
    \caption{
    The relation between pathway index and the count of closed triads by different recommendation algorithms. Comparison performed across different algorithm, existence of repost, and different memory length for opinion-based recommendation systems. All of these 3 factors could magnify the relation between the amount and the pathway index, where the larger $I_w$ , the more closed triads there will be, and vice versa.
    }
    \label{fig:supp:cont-interpret-raw}
\end{figure*}

In the main text (Figure \ref{fig:cont-interpret}), to highlight the impact of a single dimension on the social system, we averaged only that dimension. However, the effects of different dimensions can be cumulative. Therefore, we illustrate this cumulative effect in greater detail in Figure \ref{fig:supp:cont-interpret-raw}. Additionally, for opinion-based recommendation systems, longer historical memory length also lead to an increase in closed triads, as seen in Figure \ref{fig:supp:cont-interpret-raw}(e).

\section*{Potential Landscape Computation}

\subsection*{Perceived Opinion Distributions}

\paragraph{Definitions.}

For an agent at opinion $x$, let $f_n(x,x')$ denote the conditional PDF of opinions received
from followees, and $f_r(x,x')$ those from the recommendation.
Denote $\phi(x';\mu,\sigma^2)$ as the Gaussian density with mean $\mu$ and variance $\sigma^2$.
The upper/lower sign convention $(\mp\!/\!\pm)$ selects $x<0$ and $x\ge 0$.

\paragraph{Initial \& Consensual Networks.}

The idealized initial and consensual network is simply an ER network. The difference only lies on the opinion distribution. Since A pure ER network does not retain structural information, in the initial and consensual states all three distributions coincide ($f_n=f_r^{\mathrm{rand}}=f_r^{\mathrm{st}}$),
so the Random and Structure-based algorithms produce identical social forces there.

\paragraph{Diverged Network.}

The idealized diverged network is constructed by 2 random ER networks, with each agent having the followees within the same network in a proportion $r_s$, and within another network in a proportion $r_d=1-r_s$.
Therefore, the proportion that an agent with a network distance of 2 from any given agent belongs to the same or the different network is
$\tilde r_s = r_s^2+r_d^2$ and $\tilde r_d = 2r_s r_d$, respectively
. These are also the probability that a structure-based recommendation found via common neighbors lands in the same or
opposite camp, respectively.

\paragraph{Opinion-peer Distribution.}

Given the base distribution $f_r^{\mathrm{rand}}$, the opinion-peer algorithm
recommends opinions within a mass-constrained window $[l(x),u(x)]$ centered (in CDF mass) at $x$:
\begin{align}
  &f_r^{\mathrm{op}}(x,x')
  = \frac{f_r^{\mathrm{rand}}(x,x')\,\mathbf{1}_{[l(x),\,u(x)]}(x')}
         {\displaystyle\int_{l(x)}^{u(x)} f_r^{\mathrm{rand}}(x,t)\,\mathrm{d}t}, \label{eq:fop} \\ 
  &\int_{l(x)}^{u(x)} f_r^{\mathrm{rand}}(x,x')\,\mathrm{d}x' = \frac{k_r}{N-k_n}, \nonumber
\end{align}
with CDF mass drawn symmetrically from each side of $x$ (capped at domain boundaries when necessary). This serves as a mean recommendation produced by an opinion-based recommendation algorithm.

\paragraph{Conditional PDFs.}

See Table \ref{tab:supp:si-pdfs} for the definition of conditional PDFs for all cases and all sources.

\begin{table*}[h!]
\centering
\caption{Conditional opinion PDFs for each idealized network state and recommendation channel.
  $\sigma_c = \kappa_c^{-1}$, $\sigma_d = \kappa_d^{-1}$.
  The opinion-peer distribution $f_r^{\mathrm{op}}$ is defined by Eq.~\eqref{eq:fop}.}
\label{tab:supp:si-pdfs}
\renewcommand{\arraystretch}{1.7}
\setlength{\tabcolsep}{5pt}
\begin{tabular}
{>{\bfseries}
    p{2cm}
    p{3.6cm}
    p{3.6cm}
    p{3.6cm}
    p{2cm}
}
\hline
State
  & $f_n$
  & $f_r^{\mathrm{rand}}$
  & $f_r^{\mathrm{st}}$
  & $f_r^{\mathrm{op}}$ \\
\hline
Initial
  & $\tfrac{1}{2}\mathbf{1}_{[-1,1]}(x')$
  & $\tfrac{1}{2}\mathbf{1}_{[-1,1]}(x')$
  & $\tfrac{1}{2}\mathbf{1}_{[-1,1]}(x')$
  & Eq.~\eqref{eq:fop} \\
Consensual
  & $\phi(x';\,0,\sigma_c^2)$
  & $\phi(x';\,0,\sigma_c^2)$
  & $\phi(x';\,0,\sigma_c^2)$
  & Eq.~\eqref{eq:fop} \\
Diverged
  & $r_s\,\phi(x';\!\mp0.5,\sigma_d^2)$\newline
    ${}\;+r_d\,\phi(x';\!\pm0.5,\sigma_d^2)$
  & $\tfrac{1}{2}\phi(x';\!-0.5,\sigma_d^2)$\newline
    ${}\;+\tfrac{1}{2}\phi(x';\!+0.5,\sigma_d^2)$
  & $\tilde r_s\,\phi(x';\!\mp0.5,\sigma_d^2)$\newline
    ${}\;+\tilde r_d\,\phi(x';\!\pm0.5,\sigma_d^2)$
  & Eq.~\eqref{eq:fop} \\
\hline
\end{tabular}
\end{table*}

\subsection*{Derivation of the Mean-Field Social Force}

Because posts are reposts in the real model, their opinions equal the originating agent's opinion.
In the idealized analysis we therefore treat the concordant-post opinions as i.i.d.\ draws from $f_n$
and $f_r$.
Define the concordance probability and in-window conditional mean:
\begin{align*}
  P_{\epsilon,f}(x) &:= \int_{x-\epsilon}^{x+\epsilon} f(x,x')\,\mathrm{d}x', \\
  \mu_{\epsilon,f}(x) &:= \frac{\displaystyle\int_{x-\epsilon}^{x+\epsilon}x'f(x,x')\,\mathrm{d}x'}{P_{\epsilon,f}(x)}.
\end{align*}
The expected counts of concordant posts from the two channels are
$k_n P_{\epsilon,f_n}$ and $k_r P_{\epsilon,f_r}$; each concordant post contributes
a deviation $\tau-x$ whose expectation is $\mu_{\epsilon,f}-x$.
Applying a mean-field approximation (law of large numbers over the $k$ post counts), the
normalized opinion update concentrates around the social force
\begin{equation*}
  F(x)
  = \frac{k_n P_{\epsilon,f_n}(x)\,\mu_{\epsilon,f_n}(x)
         +k_r P_{\epsilon,f_r}(x)\,\mu_{\epsilon,f_r}(x)}
         {k_n P_{\epsilon,f_n}(x)+k_r P_{\epsilon,f_r}(x)}
  - x.
\end{equation*}

\section*{Numerical Results of the Potential Landscape}

To investigate the microscopic mechanisms driving opinion evolution, we analyze the trajectories of agents in the phase space of opinion $x$ and its rate of change. We introduce two metrics to quantify the ``social force'' acting upon an agent: the \emph{Normalized Opinion Difference} (NOD) and the \emph{Followees' Opinion Difference} (FOD).

\subsection*{Metric Definitions}
Let $x_i(t)$ denote the opinion of agent $i$ at time $t$. We define the NOD, denoted by $\Delta_n x_i(t)$, and the FOD, denoted by $\Delta_f x_i(t)$, as follows:
\begin{align*}
    \Delta_n x_i(t) &:= \alpha^{-1}(x_i(t + 1) - x_i(t)), \\
    \Delta_f x_i(t) &:= \frac{1}{|N_i(t)|} \sum_{(\cdot, \cdot, \tau) \in N_i(t)} (\tau - x_i(t)),
\end{align*}
where $s$ is a scaling factor for the time interval, $t_a$ represents the characteristic activation time, and $N_i(t)$ is the set of messages (or active neighbors) perceived by agent $i$ at time $t$, with $\tau$ representing the opinion value associated with a neighbor or message.

Physically, $\Delta_n x_i(t)$ represents the realized displacement of the agent's opinion over a normalized time scale, capturing the temporal tendency of the shift. Conversely, $\Delta_f x_i(t)$ represents the discrepancy between the agent's current state and the local field generated by their social neighborhood.

\subsection*{Potential Energy Estimation}
We observe that the opinion dynamics resemble the motion of an overdamped particle in a potential well, where the velocity (rate of opinion change) is proportional to the force exerted by the potential gradient. To reconstruct this underlying landscape, we treat the observed metrics as samples of a stochastic social force $F$.

Let $\mathcal{S} = \{(x_{k}, F_{k})\}_{k=1}^{M}$ be the set of all observed pairs of opinion states and corresponding forces (where $F$ corresponds to either $\Delta_n x$ or $\Delta_f x$) across all agents and relevant time steps. To estimate the macroscopic force field $\bar{F}(x)$ from these discrete, noisy observations, we employ the Nadaraya-Watson kernel regression estimator:
\begin{align*}
    \bar{F}(x) = \frac{\sum_{k=1}^{M} K_h(x - x_k) \cdot F_k}{\sum_{k=1}^{M} K_h(x - x_k)},
\end{align*}
where $K_h(u)$ is a Gaussian kernel with bandwidth $h$:
\begin{align*}
    K_h(u) = \frac{1}{\sqrt{2\pi}h} \exp\left(-\frac{u^2}{2h^2}\right).
\end{align*}
In our analysis, we set the bandwidth $h = 0.1$ to smooth local fluctuations while preserving the structural features of the distribution arising from variations in point density.

Assuming the system behaves as a gradient system where the force is the negative gradient of a potential $V(x)$ (i.e., $\bar{F}(x) = -\nabla V(x)$), we derive the potential energy landscape by integrating the smoothed force field:
\begin{align*}
    V(x) = C - \int_{-1}^{x} \bar{F}(z) \, dz.
\end{align*}
The integration is performed numerically using the trapezoidal rule. The integration constant $C$ is determined by imposing a zero-mean condition over the domain $[-1, 1]$ to ensure comparability across different time intervals:
\begin{align*}
    \int_{-1}^{1} V(x) \, dx = 0.
\end{align*}

By partitioning the simulation timeline into intervals (e.g., $t_n \in [0, 0.1), [0.1, 0.2), \dots, [0.9, \infty)$), we compute $V(x)$ for each segment. This reveals a time-varying potential landscape where the topology—specifically the depth and separation of the potential wells—evolves, characterizing the dissipation of social tension and the stability of opinion clusters.

Drawing the $(x, \Delta_nx_i)$ or $(x, \Delta_fx_i)$ trajectories w.r.t all $(i, t)$ pairs produces interesting patterns showing the microscopic mechanism of the model, as shown in Figure \ref{fig:supp:x-dx-patterns-contrast} and \ref{fig:supp:x-dx-patterns}.
This helps us to understand the patterns in Figure \ref{fig:cont-heatmap} and Figure \ref{fig:cluster-and-repost}.

\subsection*{Categorization of Microscopic Model Dynamics}

\begin{figure*}[!htbp]
    \centering
    \includegraphics[width=0.8\linewidth]{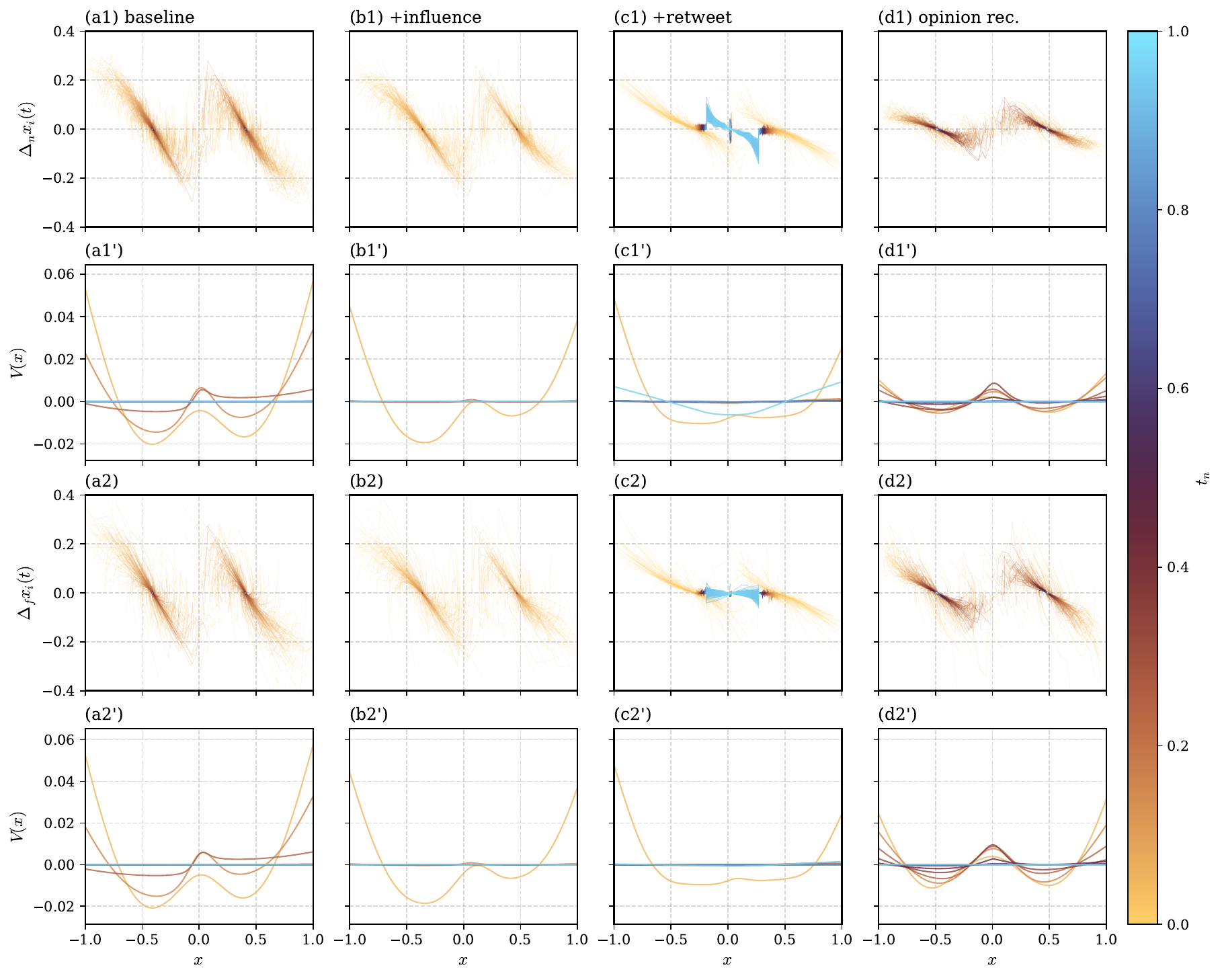}
    \caption{
    The patterns of opinion attractors emerged under (a) a baseline condition and (b--d) 3 slightly changed conditions. The first and second row shows the opinion-NOD trajectories and potential evolution, while the third and forth shows the opinion-FOD trajectories and potential evolution. The baseline is simulated under parameters $n=500, \langle k_o \rangle=15, \alpha=q=0.05, p=0$ with a random recommendation algorithm. The condition changes are: (b) $\alpha = 0.1$, (c) $p=0.1$, (d) the recommendation is shifted to an opinion-based one. The $t_a$'s are respectively (a)351, (b)441, (c)3254 and (d)605.
    }
    \label{fig:supp:x-dx-patterns-contrast}
\end{figure*}

Let us start with the patterns in Figure \ref{fig:supp:x-dx-patterns-contrast}. All of the subfigures show downward-sloping clusters with or without connections with adjacent clusters. These clusters show the direction in which an agent's opinion is attracted. An \emph{attractor} is hereafter referred to a cluster. The result shows:

\paragraph{Emergence.} The fundamental reason of an opinion attractor's formation is \emph{social influence} itself, with the mathematical form of an agent shifting its opinion to the center of all received ones, creating social forces and potential wells. As seen in Figure \ref{fig:supp:x-dx-patterns-contrast}(a), 2 attractors show on both types of trajectories, and all agents gradually cling to one of them, eventually forming a segregated network. The social force attenuates to 0 as the model evolves, forming 2 stable fixed points.

\paragraph{Factors affecting the intensity.} The attractors' intensity is hereafter determined by the strength of social influence, which corresponds to the parameter $\alpha$. In Figure \ref{fig:supp:x-dx-patterns-contrast}(b), the model shows a convergence and potential well's collapse in the earlier stage of evolution by increasing $\alpha$ from 0.05 to 0.1. Opinion-based recommendation is also capable of slowening down the convergence by creating homophily in an algorithmic basis. As shown in Figure \ref{fig:supp:x-dx-patterns-contrast}(d), while in early stages the dynamics is similar to the baseline in (a), both NOD and FOD decreases slower in later stages by stronger local homophily, causing a slower convergence.

\begin{figure*}[!htbp]
    \centering
    \includegraphics[width=0.8\linewidth]{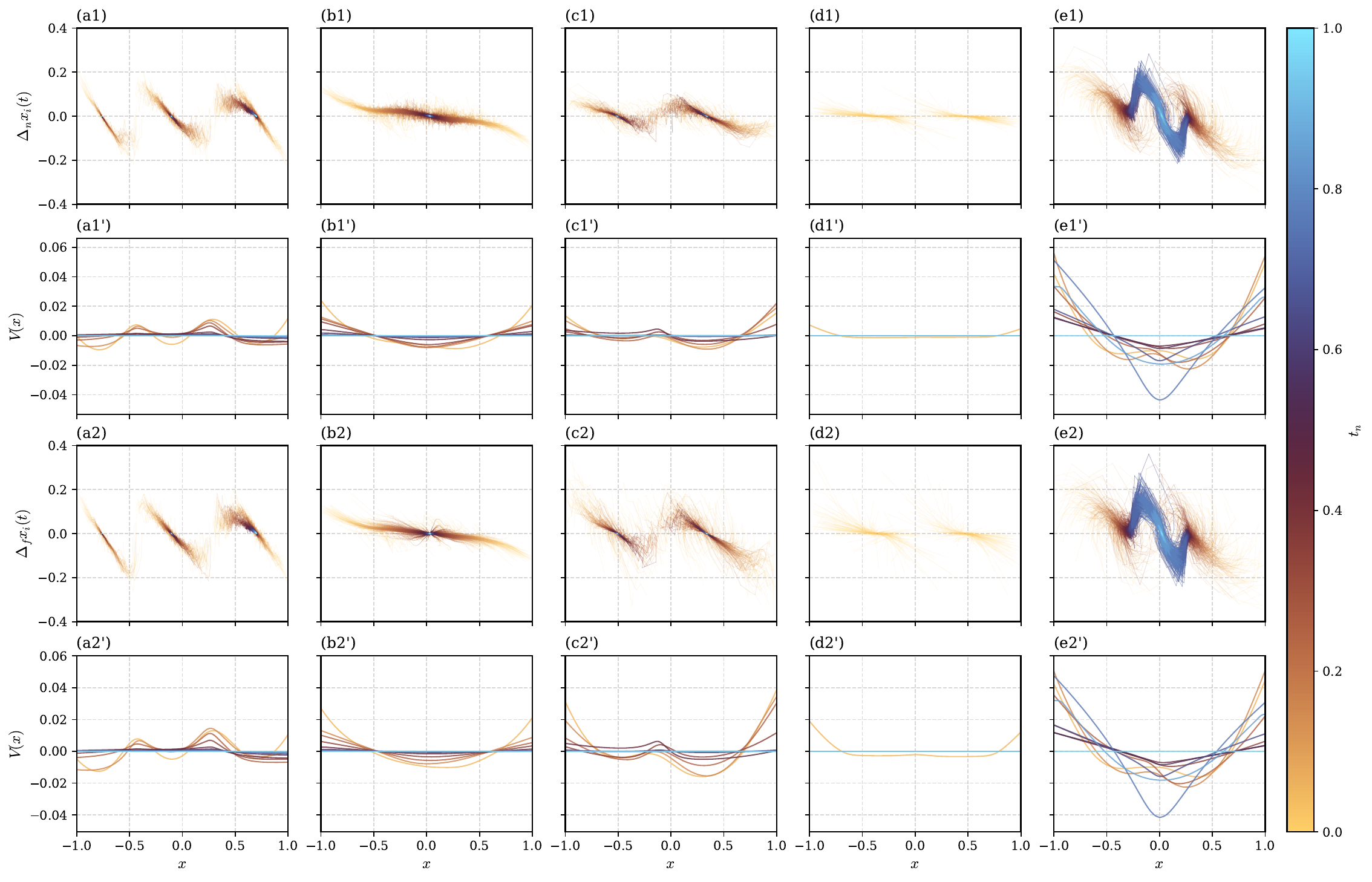}
    \caption{
    The typical patterns emerged in the trajectory diagrams, chosen from 5 simulations with parameters selected in those used in the main text.
    (a): Rewiring is too strong, directly fragmenting the network.
    (b): Rewiring is much stronger than influence, resulting in smooth influence.
    (c): Rewiring is relatively weaker than (b), an thus influence begins to become apparent.
    (d): Influence is dominant, so rewiring can act as a divisive force.
    (e): Influence is too strong, leading to abrupt large-scale opinion changes, which in turn creates consensual society.
    }
    \label{fig:supp:x-dx-patterns}
\end{figure*}

\paragraph{Abrupt merging.} In the model, influence can propagate beyond the confidence boundary by several chained agents with each one being inside the boundary of both its neighbors. Increasing the frequency of these chains' appearance causes attractors' merging. This is achieved not by setting a very large $\alpha$, but by applying either repost or structure-based recommendation. The merging is shown in Figure \ref{fig:supp:x-dx-patterns-contrast}(c) by applying repost and Figure \ref{fig:supp:x-dx-patterns}(e) by applying link-based recommendation. In both cases, the attractors are gradually approaching each other by rarely-seen cross-faction information (this also eliminates the barriers of 2 potential wells), and the influence is strong enough to allow their abrupt fusion. This also provides another implication: \emph{the only force capable of distracting an attractor is another attractor with proper opinion distance.}

\paragraph{Merging \& slicing in early stages.} Opinion-based recommendation shows the capability of both meditating or generating a diverged and segregated society in the early stages. In Figure \ref{fig:supp:x-dx-patterns-contrast}(d), although it slow down the speed of polarization, it increases the potential barrier to suppress chances of cross-faction communication, as seen in the $x_i \approx 0$ part, making the society inevitably diverged. In Figure \ref{fig:supp:x-dx-patterns}(b), a diverged society never emerged, as there is only 1 attractor at $x_i \approx 0$. In Figure \ref{fig:supp:x-dx-patterns}(a), however, with introducing reposting under same conditions of (b), the FOD becomes very large to tear the society apart to 3 attractors.

\ 

Figure \ref{fig:supp:x-dx-patterns} shows the pattern's shift under the contrast of social influence and social selection, depicted by $\alpha$ and $q$ in the model.
Let us start from Figure \ref{fig:supp:x-dx-patterns}(c). This pattern represents the most intermediate cases, where 2 attractors emerge with a relatively weaker entanglement. The result of these cases, in the balanced belt in Figure \ref{fig:cont-heatmap}, either consensual or bipolarizing, depends randomly. 
(b) and (d) shows 2 types of cases where one force is relatively stronger. 
For (b), the stronger rewiring effect enhances the entanglement of the 2 attractors, reforging them to 1, leading the system to a homogenizing pathway; In contrast, a stronger social influence breaks up the belt between 2 attractors, making them distinct and insulated, thus leading the system to a polarizing pathway. 
(a) and (e) shows 2 most extreme types where 2 forces are very unbalanced.
In (a), the effect of rewiring is so significant that 3 attractors formed; By contrast, the paramount social influence in (e) allows the abrupt merging of attractors in the late stage of evolution, similar to what seen in Figure \ref{fig:supp:x-dx-patterns-contrast}(c).
This interprets the counter-intuitive cases in Section \ref{sec:ec-pathways} and \ref{sec:ec-societal-interp}.

% \bibliographystyle{plainnat}
% \bibliography{references}

% \end{document}

    % Print references specifically for the supplement
    \printbibliography[heading=subbibliography, title={References}]

@article{degroot1974reaching,
  title   = {Reaching a Consensus},
  author  = {DeGroot, Morris H.},
  journal = {Journal of the American Statistical Association},
  volume  = {69},
  number  = {345},
  pages   = {118--121},
  year    = {1974},
  publisher = {Taylor & Francis},
  doi     = {10.1080/01621459.1974.10480137}
}

@article{deffuant2000mixing,
  title   = {Mixing Beliefs among Interacting Agents},
  author  = {Deffuant, Guillaume and Neau, David and Amblard, Fr{\'e}d{\'e}ric and Weisbuch, G{\'e}rard},
  journal = {Advances in Complex Systems},
  volume  = {3},
  number  = {1-4},
  pages   = {87--98},
  year    = {2000},
  publisher = {World Scientific},
  doi     = {10.1142/S0219525900000078}
}

@article{hegselmann2002opinion,
  title   = {Opinion Dynamics and Bounded Confidence Models, Analysis, and Simulation},
  author  = {Hegselmann, Rainer and Krause, Ulrich},
  journal = {Journal of Artificial Societies and Social Simulation},
  volume  = {5},
  number  = {3},
  pages   = {2},
  year    = {2002}
}

@inproceedings{ren2005survey,
  author    = {Ren, Wei and Beard, Randal W. and Atkins, Ella M.},
  title     = {A Survey of Consensus Problems in Multi-Agent Coordination},
  booktitle = {Proceedings of the 2005 American Control Conference},
  year      = {2005},
  pages     = {1859--1864},
  doi       = {10.1109/ACC.2005.1470239}
}

@article{sirbu2019algorithmic,
  title   = {Algorithmic Bias Amplifies Opinion Fragmentation and Polarization: A Bounded Confidence Model},
  author  = {S{\^i}rbu, Alina and Pedreschi, Dino and Giannotti, Fosca and Kert{\'e}sz, J{\'a}nos},
  journal = {PLoS ONE},
  volume  = {14},
  number  = {3},
  pages   = {e0213246},
  year    = {2019},
  publisher = {Public Library of Science},
  doi     = {10.1371/journal.pone.0213246},
  note    = {Open Access}
}

@misc{piccoli2018sparse,
  title         = {Sparse Control of Hegselmann-Krause Models: Black Hole and Declustering},
  author        = {Piccoli, Benedetto and Pouradier Duteil, Nastassia and Tr{\'e}lat, Emmanuel},
  year          = {2018},
  eprint        = {1802.00615},
  archivePrefix = {arXiv},
  primaryClass  = {math.OC},
  url           = {https://arxiv.org/abs/1802.00615}
}

@misc{pluchino2004changing,
  title         = {Changing Opinions in a Changing World: A New Perspective in Sociophysics},
  author        = {Pluchino, Alessandro and Latora, Vito and Rapisarda, Andrea},
  year          = {2004},
  eprint        = {cond-mat/0410217},
  archivePrefix = {arXiv},
  primaryClass  = {cond-mat.other},
  url           = {https://arxiv.org/abs/cond-mat/0410217}
}

@article{xue2020opinion,
  author  = {Xue, Dong and Hirche, Sandra and Cao, Ming},
  title   = {Opinion Behavior Analysis in Social Networks Under the Influence of Coopetitive Media},
  journal = {IEEE Transactions on Network Science and Engineering},
  volume  = {7},
  number  = {3},
  pages   = {961--974},
  year    = {2020},
  doi     = {10.1109/TNSE.2019.2894565}
}

@article{yan2024incorporation,
  title   = {Incorporation of Likely Future Actions of Agents into Pseudo-Gradient Dynamics of Noncooperative Games},
  author  = {Yan, Yuyue and Hayakawa, Tomohisa},
  journal = {IEEE Transactions on Automatic Control},
  volume  = {69},
  number  = {11},
  pages   = {7662--7677},
  year    = {2024},
  doi     = {10.1109/TAC.2023.3264354}
}

@article{yan2021understanding,
  title   = {Understanding How Retweets Influence the Behaviors of Social Networking Service Users via Agent-Based Simulation},
  author  = {Yan, Yizhou and Toriumi, Fujio and Sugawara, Tetsuji},
  journal = {Computational Social Networks},
  volume  = {8},
  pages   = {1--19},
  year    = {2021},
  doi     = {10.1186/s40649-021-00099-8}
}

@misc{hata2025manipulating,
  title         = {Manipulating Collective Opinion Through Social Network Intervention},
  author        = {Hata, Shigefumi and Lambiotte, Renaud and Nakao, Hiroya and Kobayashi, Ryota},
  year          = {2025},
  eprint        = {2511.12444},
  archivePrefix = {arXiv},
  primaryClass  = {physics.soc-ph},
  url           = {https://arxiv.org/abs/2511.12444}
}

@misc{larooij2025fix,
  title         = {Can We Fix Social Media? Testing Prosocial Interventions Using Generative Social Simulation},
  author        = {Larooij, Maik and T{\"o}rnberg, Petter},
  year          = {2025},
  eprint        = {2508.03385},
  archivePrefix = {arXiv},
  primaryClass  = {cs.SI},
  url           = {https://arxiv.org/abs/2508.03385}
}

@article{sasahara2021social,
  title   = {Social Influence and Unfollowing Accelerate the Emergence of Echo Chambers},
  author  = {Sasahara, Kazutoshi and Chen, Wen and Peng, Hao and Ciampaglia, Giovanni Luca and Flammini, Alessandro and Menczer, Filippo},
  journal = {Journal of Computational Social Science},
  volume  = {4},
  number  = {1},
  pages   = {381--402},
  year    = {2021},
  doi     = {10.1007/s42001-020-00084-7}
}

@inproceedings{mok2023echo,
  title     = {Echo Tunnels: Polarized News Sharing Online Runs Narrow but Deep},
  author    = {Mok, Lillio and Inzlicht, Michael and Anderson, Ashton},
  booktitle = {Proceedings of the International AAAI Conference on Web and Social Media},
  volume    = {17},
  number    = {1},
  pages     = {662--673},
  year      = {2023},
  doi       = {10.1609/icwsm.v17i1.22177},
  url       = {https://doi.org/10.1609/icwsm.v17i1.22177}
}

@article{morales2021no,
  title   = {No Echo in the Chambers of Political Interactions on Reddit},
  author  = {De Francisci Morales, Gianmarco and Monti, Corrado and Starnini, Michele},
  journal = {Scientific Reports},
  volume  = {11},
  pages   = {2818},
  year    = {2021},
  doi     = {10.1038/s41598-021-81531-x}
}

@article{bail2018exposure,
  title   = {Exposure to Opposing Views on Social Media Can Increase Political Polarization},
  author  = {Bail, Christopher A. and Argyle, Lisa P. and Brown, Taylor W. and Bumpus, John P. and Chen, Haohan and Fallin Hunzaker, M. B. and Lee, Jaemin and Mann, Marcus and Merhout, Friedolin and Volfovsky, Alexander},
  journal = {Proceedings of the National Academy of Sciences},
  volume  = {115},
  number  = {37},
  pages   = {9216--9221},
  year    = {2018},
  doi     = {10.1073/pnas.1804840115}
}

@article{grover2019polarization,
  title   = {Polarization and Acculturation in US Election 2016 Outcomes: Can Twitter Analytics Predict Changes in Voting Preferences?},
  author  = {Grover, Purva and Kar, Arpan and Dwivedi, Yogesh K. and Janssen, Marijn},
  journal = {Technological Forecasting and Social Change},
  volume  = {145},
  pages   = {438--452},
  year    = {2019},
  doi     = {10.1016/j.techfore.2018.09.009}
}

@article{huszar2022algorithmic,
  author  = {Husz{\'a}r, Ferenc and Ktena, Sofia Ira and O'Brien, Conor and Belli, Luca and Schlaikjer, Andrew and Hardt, Moritz},
  title   = {Algorithmic Amplification of Politics on Twitter},
  journal = {Proceedings of the National Academy of Sciences},
  volume  = {119},
  number  = {1},
  pages   = {e2025334119},
  year    = {2022},
  doi     = {10.1073/pnas.2025334119}
}

@article{buntain2021youtube,
  author  = {Buntain, Cody and Bonneau, Richard and Nagler, Jonathan and Tucker, Joshua A.},
  title   = {YouTube Recommendations and Effects on Sharing Across Online Social Platforms},
  journal = {Proceedings of the ACM on Human-Computer Interaction},
  volume  = {5},
  number  = {CSCW1},
  pages   = {11:1--11:26},
  year    = {2021},
  doi     = {10.1145/3449085}
}

@article{liu2023daily,
  title   = {The Daily Me Versus the Daily Others: How Do Recommendation Algorithms Change User Interests? Evidence from a Knowledge-Sharing Platform},
  author  = {Liu, Jia and Cong, Ziwei},
  journal = {Journal of Marketing Research},
  volume  = {60},
  number  = {4},
  pages   = {767--791},
  year    = {2023},
  doi     = {10.1177/00222437221134237}
}

@article{jasny2015empirical,
  author  = {Jasny, Lorien and Waggle, Joseph and Fisher, Dana R.},
  title   = {An Empirical Examination of Echo Chambers in US Climate Policy Networks},
  journal = {Nature Climate Change},
  volume  = {5},
  pages   = {782--786},
  year    = {2015},
  doi     = {10.1038/nclimate2666}
}

@article{kaiser2020birds,
  author  = {Kaiser, Jonas and Rauchfleisch, Adrian},
  title   = {Birds of a Feather Get Recommended Together: Algorithmic Homophily in YouTube’s Channel Recommendations in the United States and Germany},
  journal = {Social Media + Society},
  volume  = {6},
  number  = {4},
  pages   = {1--15},
  year    = {2020},
  doi     = {10.1177/2056305120969914}
}

@article{bozdag2020managing,
  author  = {Bozdag, Cigdem},
  title   = {Managing Diverse Online Networks in the Context of Polarization: Understanding How We Grow Apart on and through Social Media},
  journal = {Social Media + Society},
  volume  = {6},
  number  = {4},
  pages   = {2056305120975713},
  year    = {2020},
  doi     = {10.1177/2056305120975713}
}

@article{santos2021link,
  title   = {Link Recommendation Algorithms and Dynamics of Polarization in Online Social Networks},
  author  = {Santos, Fernando P. and Lelkes, Yphtach and Levin, Simon A.},
  journal = {Proceedings of the National Academy of Sciences},
  volume  = {118},
  number  = {50},
  pages   = {e2102141118},
  year    = {2021},
  doi     = {10.1073/pnas.2102141118}
}

@article{tornberg2022digital,
  title   = {How Digital Media Drive Affective Polarization through Partisan Sorting},
  author  = {T{\"o}rnberg, Petter},
  journal = {Proceedings of the National Academy of Sciences},
  volume  = {119},
  number  = {42},
  pages   = {e2207159119},
  year    = {2022},
  doi     = {10.1073/pnas.2207159119}
}

@misc{alatawi2021survey,
  title         = {A Survey on Echo Chambers on Social Media: Description, Detection and Mitigation},
  author        = {Alatawi, Faisal and Cheng, Lu and Tahir, Anique and Karami, Mansooreh and Jiang, Bohan and Black, Tyler and Liu, Huan},
  year          = {2021},
  eprint        = {2112.05084},
  archivePrefix = {arXiv},
  primaryClass  = {cs.SI},
  url           = {https://arxiv.org/abs/2112.05084}
}

@article{dearruda2024echo,
  title   = {Echo Chamber Formation Sharpened by Priority Users},
  author  = {de Arruda, Henrique F. and Oliveira, Kleber A. and Moreno, Yamir},
  journal = {iScience},
  volume  = {27},
  number  = {11},
  pages   = {111098},
  year    = {2024},
  doi     = {10.1016/j.isci.2024.111098},
  note    = {Preprint available as arXiv:2312.09358}
}

@misc{duskin2024echo,
  title         = {Echo Chambers in the Age of Algorithms: An Audit of Twitter's Friend Recommender System},
  author        = {Duskin, Kayla and Schafer, Joseph S. and West, Jevin D. and Spiro, Emma S.},
  year          = {2024},
  eprint        = {2404.06422},
  archivePrefix = {arXiv},
  primaryClass  = {cs.SI},
  url           = {https://arxiv.org/abs/2404.06422}
}

@inproceedings{morales2021auditing,
  author    = {Ramaciotti Morales, Pedro and Cointet, Jean-Philippe},
  title     = {Auditing the Effect of Social Network Recommendations on Polarization in Geometrical Ideological Spaces},
  booktitle = {Proceedings of the 15th ACM Conference on Recommender Systems},
  year      = {2021},
  pages     = {627--632},
  doi       = {10.1145/3460231.3478851}
}

@inproceedings{jiang2019degenerate,
  series    = {AIES '19},
  title     = {Degenerate Feedback Loops in Recommender Systems},
  author    = {Jiang, Ray and Chiappa, Silvia and Lattimore, Tor and Gy{\"o}rgy, Andr{\'a}s and Kohli, Pushmeet},
  booktitle = {Proceedings of the 2019 AAAI/ACM Conference on AI, Ethics, and Society},
  year      = {2019},
  pages     = {383--390},
  publisher = {ACM},
  doi       = {10.1145/3306618.3314288},
  url       = {https://doi.org/10.1145/3306618.3314288}
}

@inproceedings{kalimeris2021preference,
  author    = {Kalimeris, Dimitris and Bhagat, Smriti and Kalyanaraman, Shankar and Weinsberg, Udi},
  title     = {Preference Amplification in Recommender Systems},
  booktitle = {Proceedings of the 27th ACM SIGKDD Conference on Knowledge Discovery and Data Mining},
  year      = {2021},
  pages     = {805--815},
  publisher = {ACM},
  doi       = {10.1145/3447548.3467298}
}

@inproceedings{schmit2018human,
  title     = {Human Interaction with Recommendation Systems},
  author    = {Schmit, Sven and Riquelme, Carlos},
  booktitle = {Proceedings of the Twenty-First International Conference on Artificial Intelligence and Statistics},
  series    = {Proceedings of Machine Learning Research},
  volume    = {84},
  pages     = {862--870},
  year      = {2018},
  publisher = {PMLR},
  url       = {https://proceedings.mlr.press/v84/schmit18a.html}
}

@inproceedings{dean2022preference,
  title     = {Preference Dynamics Under Personalized Recommendations},
  author    = {Dean, Sarah and Morgenstern, Jamie},
  booktitle = {Proceedings of the 23rd ACM Conference on Economics and Computation},
  series    = {EC '22},
  year      = {2022},
  pages     = {563--583},
  publisher = {ACM},
  doi       = {10.1145/3490486.3538346}
}

@inproceedings{chaney2018algorithmic,
  author    = {Chaney, Allison J. B. and Stewart, Brandon M. and Engelhardt, Barbara E.},
  title     = {How Algorithmic Confounding in Recommendation Systems Increases Homogeneity and Decreases Utility},
  booktitle = {Proceedings of the 12th ACM Conference on Recommender Systems},
  year      = {2018},
  pages     = {224--232},
  publisher = {ACM},
  doi       = {10.1145/3240323.3240370}
}

@inproceedings{mansoury2020feedback,
  author    = {Mansoury, Masoud and Abdollahpouri, Himan and Pechenizkiy, Mykola and Mobasher, Bamshad and Burke, Robin},
  title     = {Feedback Loop and Bias Amplification in Recommender Systems},
  booktitle = {Proceedings of the 29th ACM International Conference on Information \& Knowledge Management},
  year      = {2020},
  pages     = {2145--2148},
  publisher = {ACM},
  doi       = {10.1145/3340531.3412152}
}

@inproceedings{anderson2020algorithmic,
  author    = {Anderson, Ashton and Maystre, Lucas and Anderson, Ian and Mehrotra, Rishabh and Lalmas, Mounia},
  title     = {Algorithmic Effects on the Diversity of Consumption on Spotify},
  booktitle = {Proceedings of The Web Conference 2020},
  year      = {2020},
  pages     = {2155--2165},
  publisher = {ACM},
  doi       = {10.1145/3366423.3380281}
}

@inproceedings{donkers2021dual,
  author    = {Donkers, Tim and Ziegler, J{\"u}rgen},
  title     = {The Dual Echo Chamber: Modeling Social Media Polarization for Interventional Recommending},
  booktitle = {Proceedings of the 15th ACM Conference on Recommender Systems},
  year      = {2021},
  pages     = {12--22},
  publisher = {ACM},
  doi       = {10.1145/3460231.3474261}
}

@article{li2024recent,
  title   = {Recent Developments in Recommender Systems: A Survey},
  author  = {Li, Yang and Liu, Kangbo and Satapathy, Ranjan and Wang, Suhang and Cambria, Erik},
  journal = {IEEE Computational Intelligence Magazine},
  volume  = {19},
  number  = {2},
  pages   = {78--95},
  year    = {2024},
  doi     = {10.1109/MCI.2024.3363984},
  note    = {Review article; preprint available as arXiv:2306.12680}
}

@inproceedings{hassan2019trust,
  author    = {Hassan, Taha},
  title     = {Trust and Trustworthiness in Social Recommender Systems},
  booktitle = {Companion Proceedings of The 2019 World Wide Web Conference},
  year      = {2019},
  pages     = {529--532},
  publisher = {ACM},
  doi       = {10.1145/3308560.3317596}
}

@inproceedings{grewal2018evolution,
  author    = {Grewal, Ajeet and Lin, Jimmy},
  title     = {The Evolution of Content Analysis for Personalized Recommendations at Twitter},
  booktitle = {Proceedings of the 41st International ACM SIGIR Conference on Research and Development in Information Retrieval},
  year      = {2018},
  pages     = {1355--1356},
  publisher = {ACM},
  doi       = {10.1145/3209978.3210206}
}

@inproceedings{ying2022recommendation,
  author    = {Ying, HongDa and Takano, Kosuke},
  title     = {A Recommendation Method for Social Media Users Based on a Sentiment Analysis Model},
  booktitle = {2022 IEEE 4th Global Conference on Life Sciences and Technologies (LifeTech)},
  year      = {2022},
  pages     = {485--488},
  publisher = {IEEE},
  doi       = {10.1109/LifeTech53646.2022.9754863}
}

@inproceedings{yang2022personality,
  author    = {Yang, Qi and Nikolenko, Sergey and Huang, Alfred and Farseev, Aleksandr},
  title     = {Personality-Driven Social Multimedia Content Recommendation},
  booktitle = {Proceedings of the 30th ACM International Conference on Multimedia},
  series    = {MM '22},
  year      = {2022},
  publisher = {ACM},
  doi       = {10.1145/3503161.3548769}
}

@inproceedings{xiao2019beyond,
  title     = {Beyond Personalization: Social Content Recommendation for Creator Equality and Consumer Satisfaction},
  author    = {Xiao, Wenyi and Zhao, Huan and Pan, Haojie and Song, Yangqiu and Zheng, Vincent W. and Yang, Qiang},
  booktitle = {Proceedings of the 25th ACM SIGKDD International Conference on Knowledge Discovery \& Data Mining},
  series    = {KDD '19},
  year      = {2019},
  pages     = {235--245},
  publisher = {ACM},
  doi       = {10.1145/3292500.3330965}
}

@article{rossi2022closed,
  author   = {Rossi, Wilbert Samuel and Polderman, Jan Willem and Frasca, Paolo},
  title    = {The Closed Loop Between Opinion Formation and Personalized Recommendations},
  journal  = {IEEE Transactions on Control of Network Systems},
  volume   = {9},
  number   = {3},
  pages    = {1092--1103},
  year     = {2022},
  doi      = {10.1109/TCNS.2021.3105616}
}

@misc{gao2023s3,
  title         = {S3: Social-network Simulation System with Large Language Model-Empowered Agents},
  author        = {Gao, Chen and Lan, Xiaochong and Lu, Zhihong and Mao, Jinzhu and Piao, Jinghua and Wang, Huandong and Jin, Depeng and Li, Yong},
  year          = {2023},
  eprint        = {2307.14984},
  archivePrefix = {arXiv},
  primaryClass  = {cs.SI},
  url           = {https://arxiv.org/abs/2307.14984}
}

@article{ziems2024large,
  title   = {Can Large Language Models Transform Computational Social Science?},
  author  = {Ziems, Caleb and Held, William and Shaikh, Omar and Chen, Jiaao and Zhang, Zhehao and Yang, Diyi},
  journal = {Computational Linguistics},
  volume  = {50},
  number  = {1},
  pages   = {237--291},
  year    = {2024},
  doi     = {10.1162/coli_a_00502},
  note    = {Preprint available as arXiv:2305.03514}
}

@article{lin1991divergence,
  title   = {Divergence Measures Based on the Shannon Entropy},
  author  = {Lin, Jianhua},
  journal = {IEEE Transactions on Information Theory},
  volume  = {37},
  number  = {1},
  pages   = {145--151},
  year    = {1991},
  doi     = {10.1109/18.61115}
}

@article{zignani2014link, 
title={Link and Triadic Closure Delay: Temporal Metrics for Social Network Dynamics}, 
volume={8}, 
url={https://ojs.aaai.org/index.php/ICWSM/article/view/14507}, 
doi={10.1609/icwsm.v8i1.14507}, 
number={1}, 
journal={Proceedings of the International AAAI Conference on Web and Social Media}, 
author={Zignani, Matteo and Gaito, Sabrina and Rossi, Gian Paolo and Zhao, Xiaohan and Zheng, Haitao and Zhao, Ben}, 
year={2014}, 
pages={564-573} 
}

@article{friedkin1990social,
  author  = {Noah E. Friedkin and Eugene C. Johnsen},
  title   = {Social Influence and Opinions},
  journal = {The Journal of Mathematical Sociology},
  volume  = {15},
  number  = {3-4},
  pages   = {193--206},
  year    = {1990},
  doi     = {10.1080/0022250X.1990.9990069}
}

@article{friedkin1999social,
  author  = {Noah E. Friedkin and Eugene C. Johnsen},
  title   = {Network Structure and Influence Satisfaction},
  journal = {Advances in Group Processes},
  volume  = {16},
  pages   = {1--37},
  year    = {1999}
}

@article{jannach2017efficient,
  author  = {Dietmar Jannach and Lukas Lerche and Iman Kamehkhosh and Michael Jugovac},
  title   = {Efficient optimization of multiple recommendation quality factors according to individual user preferences},
  journal = {Expert Systems with Applications},
  volume  = {81},
  pages   = {321--331},
  year    = {2017},
  doi     = {10.1016/j.eswa.2017.03.061}
}

@inproceedings{ekstrand2018all,
  author    = {Michael D. Ekstrand and Mohammed Imran and Mucun Tian and Raghavendra Uppu and Daniel Kluver and Joseph A. Konstan and John T. Riedl},
  title     = {All the Cool Kids, How Do They Fit In?: Popularity and Demographic Biases in Recommender System Evaluation and Effectiveness},
  booktitle = {Proceedings of the 1st Conference on Fairness, Accountability and Transparency},
  volume    = {81},
  pages     = {1--15},
  year      = {2018},
  publisher = {PMLR}
}

@article{wu2022survey,
  author  = {Chunlin Wu and Yuming Li and Yanzhe Wang and Yifan Liu},
  title   = {A Survey on Personality-Aware Recommendation Systems},
  journal = {Systems},
  volume  = {10},
  number  = {6},
  pages   = {198},
  year    = {2022},
  doi     = {10.3390/systems10060198}
}

@inproceedings{ren2014random,
  author    = {Xiang Ren and Jialie Shen and Xiao Bai},
  title     = {Random Walk-based Recommendation with Restart using Social Information and Bayesian Transition Matrices},
  booktitle = {Proceedings of the 23rd International Conference on World Wide Web},
  pages     = {99--100},
  year      = {2014},
  doi       = {10.1145/2567948.2577302}
}

@article{galam2023unanimity,
  author  = {Serge Galam},
  title   = {Unanimity, Coexistence, and Rigidity: Three Sides of Polarization},
  journal = {Entropy},
  volume  = {25},
  number  = {4},
  pages   = {622},
  year    = {2023},
  doi     = {10.3390/e25040622}
}

@article{oliveira2026mechanistic,
  author  = {Kleber Andrade Oliveira and Henrique Ferraz de Arruda and Yamir Moreno},
  title   = {Mechanistic interplay between information spreading and opinion polarization},
  journal = {PNAS Nexus},
  volume  = {5},
  number  = {1},
  pages   = {pgaf402},
  year    = {2026},
  doi     = {10.1093/pnasnexus/pgaf402}
}

@Article{galam2024spontaneous,
AUTHOR = {Galam, Serge},
TITLE = {Spontaneous Symmetry Breaking, Group Decision-Making, and Beyond: 1. Echo Chambers and Random Polarization},
JOURNAL = {Symmetry},
VOLUME = {16},
YEAR = {2024},
NUMBER = {12},
ARTICLE-NUMBER = {1566},
URL = {https://www.mdpi.com/2073-8994/16/12/1566},
ISSN = {2073-8994},
DOI = {10.3390/sym16121566}
}

@article{lorenzspreen2019accelerating,
  author  = {Lorenz-Spreen, Philipp and M{\o}nsted, Bjarke M. and H{\"o}vel, Philipp and Lehmann, Sune},
  title   = {Accelerating Dynamics of Collective Attention},
  journal = {Nature Communications},
  year    = {2019},
  volume  = {10},
  number  = {1},
  pages   = {1759},
  doi     = {10.1038/s41467-019-09311-w}
}

@inproceedings{conover2011political,
  author    = {Conover, Michael D. and Ratkiewicz, Jacob and Francisco, Matthew and Gon{\c{c}}alves, Bruno and Flammini, Alessandro and Menczer, Filippo},
  title     = {Political Polarization on Twitter},
  booktitle = {Proceedings of the Fifth International AAAI Conference on Weblogs and Social Media},
  series    = {ICWSM},
  year      = {2011},
  pages     = {89--96},
  publisher = {AAAI Press},
  doi       = {10.1609/icwsm.v5i1.14126}
}

@article{fletcher2018are,
  author  = {Fletcher, Richard and Nielsen, Rasmus Kleis},
  title   = {Are People Incidentally Exposed to News on Social Media? A Comparative Analysis},
  journal = {New Media \& Society},
  year    = {2018},
  volume  = {20},
  number  = {7},
  pages   = {2450--2468},
  doi     = {10.1177/1461444817724170}
}

@article{fletcher2018automated,
  author  = {Fletcher, Richard and Nielsen, Rasmus Kleis},
  title   = {Automated Serendipity: The Effect of Using Search Engines on News Repertoire Balance and Diversity},
  journal = {Digital Journalism},
  year    = {2018},
  volume  = {6},
  number  = {8},
  pages   = {976--989},
  doi     = {10.1080/21670811.2018.1502045}
}

@inproceedings{garimella2018political,
  author    = {Garimella, Kiran and De Francisci Morales, Gianmarco and Gionis, Aristides and Mathioudakis, Michael},
  title     = {Political Discourse on Social Media: Echo Chambers, Gatekeepers, and the Price of Bipartisanship},
  booktitle = {Proceedings of the 2018 World Wide Web Conference},
  series    = {WWW '18},
  year      = {2018},
  pages     = {913--922},
  publisher = {International World Wide Web Conferences Steering Committee},
  doi       = {10.1145/3178876.3186139}
}

\end{refsection}

\end{document}